\documentclass{JHEP3}
\usepackage{graphics}
\usepackage{epsfig}
\usepackage{amsmath}
\usepackage{bbm}
\usepackage{subfigure}
\usepackage{verbatim}
\usepackage{cite}

\title{Super-leading logarithms in non-global observables in QCD:
Colour basis independent calculation}
\author{J.R. Forshaw, A. Kyrieleis \\School of Physics \& Astronomy,
University of Manchester, \\
Oxford Road, Manchester M13 9PL, U.K.\\
\email{forshaw@mail.cern.ch}, \email{kyrieleis@hep.man.ac.uk}}
\author{M.H. Seymour \\ School of Physics \& Astronomy, University of
Manchester and \\
Theoretical Physics Group (PH-TH), CERN, CH-1211 Geneva 23, Switzerland
\\ \email{mike.seymour@cern.ch}}
\preprint{MAN/HEP/2008/12 \\ CERN--PH--TH/2008--163}
\abstract{In a previous paper we reported the discovery of super-leading
logarithmic terms in a non-global QCD observable. In this short update we
recalculate the first super-leading logarithmic contribution to the
`gaps between jets' cross-section using a colour basis independent
notation.  This sheds light on the structure and origin of
the super-leading terms and allows them to be calculated for gluon
scattering processes for the first time.}
\begin{document}

\section{Introduction}

In perturbative calculations of the cross-section for the production of a pair
of jets with a rapidity gap between them, it is often assumed that the observable 
is fully inclusive outside the gap region and therefore that there is a perfect real--virtual
cancellation there \cite{Kidonakis:1998nf}--\cite{Forshaw:2005sx}.  In \cite{Forshaw:2006fk} we made a calculation of the first correction to this picture coming from one (real or virtual) gluon being emitted outside the gap and dressed by an arbitrary number of additional virtual gluons\footnote{We explain more precisely what we mean by a virtual gluon being `inside' or `outside' the gap
below.}. Based on the work of \cite{DasSa1}--\cite{Appleby:2002ke}, we
expected to find an additional tower of leading logarithms, known as
non-global logarithms, generated by the fact that gluons outside the gap
are prevented from radiating into the gap region by the gap requirement.  At leading
order this is an edge effect: gluons just outside the gap are suppressed
by the non-emission just inside the gap, leading to the existence of a `buffer
zone' in all-orders calculations \cite{DasSa2}.

However, we were surprised to find an additional long-range source of
mis-cancellation that leads to a tower of \emph{super-leading\/}
logarithms.  We traced their origin to the imaginary parts of the loop
integrals, sometimes known as the Coulomb gluon contributions.  We view this as being due to a
breakdown of naive coherence for initial-state
radiation \cite{Kyrieleis:2006fh,Seymour:2007vw}.  A similar
conclusion was also reached, for a different process, in \cite{Collins:2007nk,Collins:2007jp}.

Although we expect that our conclusions are valid for any QCD scattering
process, our calculation was only actually performed for the specific
case of quark--quark scattering, $qq\to qq$.  Generalizing this to other
$2\to2$ scattering processes, let alone the general $2\to n$ case, is
troublesome due to the large dimensionality of the colour bases in which
the anomalous dimension matrices need to be calculated.  Diagrammatic
approaches \cite{James} do not require an explicit colour basis and do
not get any more complicated when replacing the quarks by gluons, but
their disadvantage is that the number of cut diagrams soon becomes
prohibitively large.

Historically, in discussions that may turn out to be quite relevant to
the present case \cite{Catani:1985xt}, a significant advantage was
achieved by performing the calculation in a colour basis
independent way.  The purpose of the present paper is to repeat the
calculation of \cite{Forshaw:2006fk} in the colour basis
independent notation.  Of course the results agree with those of \cite{Forshaw:2006fk}, but they are obtained more transparently,
and in a way that generalizes easily to other processes.  We therefore
believe that they add considerably to the understanding of the origin
and structure of the super-leading logarithms, although they do nothing
to solve the problem of how to deal with them in general.

We view the present paper as an addendum to \cite{Forshaw:2006fk},
and therefore do not repeat the introduction, motivation or discussion
of numerical results that are contained there.  We begin in
Section \ref{summary} with a brief summary of the previous calculations,
before introducing the colour basis independent notation in
Section \ref{colour}.  In Section \ref{onegluon} we use it to recalculate
the leading non-zero contribution to the `one gluon outside the gap'
cross-section and show that it boils down to
the sum over a relatively small number of Feynman diagrams.
The method also enables us to calculate, for the first time, the coefficient of the first super-leading
logarithm for gluon scattering processes and this is done in Section~\ref{results}.
Finally, in Section~\ref{outlook}, we
make a brief outlook.  Sections~\ref{onegluon} and~\ref{results} contain
new results, while Sections~\ref{summary} and~\ref{colour} are
important to provide a pedagogical introduction to the calculation.

\section{Summary of previous calculations}
\label{summary}

The main non-trivial aspect when calculating gap cross-sections in hadron
collisions is the fact that the hard scattering matrix elements have a
non-trivial colour structure.  The simplest case, $qq\to qq$, for
example, has two independent colour structures.  These form a vector
space and for a concrete calculation we can introduce a basis for this
space.  In this example, we take the $t$-channel basis in which the two
colour structures correspond to the exchange of a singlet or an octet in
the $t$-channel.  In order to anticipate the structure of the colour
basis independent calculation, we choose to normalize our colour states,
which will mean that the soft matrix $\mathbf{S}_V$, defined in \cite{Forshaw:2006fk}, is equal to the identity matrix.  This is the only difference in notation between this section and \cite{Forshaw:2006fk}.  With this normalization, the colour basis states for $q_iq_j\to q_kq_l$ are
\begin{eqnarray}
  \mathbf{C}_1 &\equiv& \frac1N \; \delta_{ki}\,\delta_{lj}\,, \\
  \mathbf{C}_8 &\equiv& \frac2{\sqrt{N^2-1}} \; T_{ki}^a\,T_{lj}^a\,, \label{eq:4qbasis}
\end{eqnarray}
where $N$ is the number of colours.

In general, to make an all-orders resummed calculation in this process, one needs to calculate the contribution from an arbitrary number of additional virtual or real gluons.  Virtual
gluons do not change the dimensionality of the colour space, but real
gluons do: each emitted gluon takes us to a higher-dimensional space, so
the fully general all-orders calculation becomes intractable.  However,
one can make a series of steps to reformulate the all-orders calculation
as an evolution of the original $2\to2$ process in colour space. We start by outlining the thinking underlying the original calculations.
\begin{enumerate}
\item
To extract the leading logarithms of $Q/Q_0$ ($Q$ is the hard scattering scale and $Q_0$ is the veto scale used to define the gap) one can make a strong
ordering approximation.  The calculation of the $n$-loop correction to
the $2\to2$ process therefore nests and one can apply the 1-loop
correction, with the loop gluon attached only to the external partons,
$n$ times, giving an exponentiating structure.
\item
One can reduce the dimensionality of the loop integral by one by integrating over (e.g.) the
energy. The corresponding contour integral contains two kinds of pole: one where the exchanged gluon
goes on shell (which we refer to as the eikonal gluon contribution) and one
where the external partons go on shell (which we refer to as the Coulomb
gluon contribution).  The former has exactly the same structure as the
phase-space integral for emission of a real gluon; the latter does not.
\item
\label{suspect}
It is easy to show that for a fully-inclusive observable, the eikonal
gluon contribution is exactly equal and opposite to the real gluon
emission one.  In conventional calculations, it is assumed that
the observable is sufficiently inclusive for this cancellation to be
maintained for all emissions  `outside the gap'.
\item
As a result of step \ref{suspect}, real gluon emission does not contribute since outside
the gap (and below threshold within the gap) we have assumed a perfect
real--virtual cancellation whilst inside the gap and above threshold, real
emission is forbidden since it spoils the gap definition.
\end{enumerate}
Under these four assumptions, the all-orders calculation involves calculating
only virtual gluon corrections with the eikonal gluons integrated over the gap region. One finds that
\begin{equation}
  \sigma = \mathbf{M}^\dagger \, \mathbf{M},
\end{equation}
with
\begin{equation}
  \mathbf{M} = \left(\begin{array}{c}M^{(1)}\\M^{(8)}\end{array}\right).
\end{equation}
The colour evolution is given by
\begin{equation}
  \mathbf{M}(Q_0)=\exp\left(-\frac{2\alpha_s}{\pi}
  \int_{Q_0}^Q\frac{dk_T}{k_T}\,\mathbf{\Gamma}\right)
  \mathbf{M}(Q), \label{eq:expo}
\end{equation}
with boundary condition
\begin{equation}
  \mathbf{M}(Q) \equiv \mathbf{M}_0 =
  \left(\begin{array}{c}0\\\sqrt{\sigma_{\mathrm{born}}}\end{array}\right)
\end{equation}
and anomalous dimension matrix
\begin{equation}
  \mathbf{\Gamma} =
  \left(\begin{array}{cc}
    \frac{N^2-1}{4N}\rho(Y,\Delta y) & \frac{\sqrt{N^2-1}}{2N}i\pi \\
    \frac{\sqrt{N^2-1}}{2N}i\pi & -\frac1Ni\pi + \frac N2Y
    +\frac{N^2-1}{4N}\rho(Y,\Delta y)\end{array}\right).
\end{equation}
Note that, as pointed out in \cite{Seymour:2005ze} and proved in
\cite{Malin}, $\mathbf{\Gamma}$ is symmetric in this, normalized,
basis.  The function $\rho$ appearing here is defined in
\cite{Forshaw:2006fk}: it is small in the region of interest (large $Y$) and
not very relevant to the present discussion. Note that (\ref{eq:expo}) includes the $i \pi$ terms generated by Coulomb gluon emissions only at $k_T > Q_0$. The contribution from these virtual corrections
below $Q_0$ does not cancel but instead exponentiates to produce a net phase
in the amplitude that does not contribute to the cross-section.

The aim of the present calculation is to check the validity of the assumption
articulated in step~\ref{suspect} above, by calculating the correction coming
from allowing one gluon outside the gap, $\sigma_1$.  This can be written as the
sum of a real contribution plus an eikonal virtual contribution\footnote{It makes no sense to speak of Coulomb gluons being in or out of the gap.}, each
integrated over the phase space region outside the gap and each dressed
with any number of  Coulomb gluons or in-gap eikonal gluons. Thus
\begin{equation}
  \label{master}
  \sigma_1 = -\frac{2\alpha_s}{\pi}\int_{Q_0}^Q\frac{dk_T}{k_T}
  \int_{\mathrm{out}}\frac{dy\,d\phi}{2\pi}
  \Bigl(\Omega_R+\Omega_V\Bigr),
\end{equation}
where
\begin{eqnarray}
  \label{masterV}
  \Omega_V &=& \mathbf{M}_0^\dagger
  \exp\left(-\frac{2\alpha_s}{\pi}\int_{Q_0}^Q\frac{dk_T'}{k_T'}
    \mathbf{\Gamma}^\dagger\right)
\nonumber\\&&\hspace*{7.5em}
  \exp\left(-\frac{2\alpha_s}{\pi}\int_{Q_0}^{k_T}\frac{dk_T'}{k_T'}
    \mathbf{\Gamma}\right)
  \boldsymbol{\gamma}
  \exp\left(-\frac{2\alpha_s}{\pi}\int_{k_T}^Q\frac{dk_T'}{k_T'}
    \mathbf{\Gamma}\right)
  \mathbf{M}_0+\mathrm{c.c.},\phantom{(9.9)} \\
  \label{masterR}
  \Omega_R &=& \mathbf{M}_0^\dagger
  \exp\left(-\frac{2\alpha_s}{\pi}\int_{k_T}^Q\frac{dk_T'}{k_T'}
    \mathbf{\Gamma}^\dagger\right)
  \mathbf{D}_\mu^\dagger
  \exp\left(-\frac{2\alpha_s}{\pi}\int_{Q_0}^{k_T}\frac{dk_T'}{k_T'}
    \mathbf{\Lambda}^\dagger\right)
\nonumber\\&&\hspace*{7.5em}
  \exp\left(-\frac{2\alpha_s}{\pi}\int_{Q_0}^{k_T}\frac{dk_T'}{k_T'}
    \mathbf{\Lambda}\right)
  \mathbf{D}^\mu
  \exp\left(-\frac{2\alpha_s}{\pi}\int_{k_T}^Q\frac{dk_T'}{k_T'}
    \mathbf{\Gamma}\right)
  \mathbf{M}_0.
\end{eqnarray}
The two contributions have a common evolution from $Q$ down to $k_T$,
followed by, for the virtual contribution, a virtual eikonal emission at
scale $k_T$,
\begin{equation}
  \boldsymbol{\gamma} = \frac12\left(
  \begin{array}{cc}
    \frac{N^2-1}{2N}(\omega_{13}+\omega_{24}) &
    \frac{\sqrt{N^2-1}}{2N}(-\omega_{12}-\omega_{34}+\omega_{14}+\omega_{23})
    \\
    \frac{\sqrt{N^2-1}}{2N}(-\omega_{12}-\omega_{34}+\omega_{14}+\omega_{23})&
    \begin{array}[t]{c}
      \frac{N}{2}(\omega_{14}+\omega_{23})
      -\frac1{2N}(\omega_{13}+\omega_{24}) \\
      +\frac1N(\omega_{12}+\omega_{34}-\omega_{14}-\omega_{23})
    \end{array}\end{array}\right),
\end{equation}
which can appear on either side of the cut, with
\begin{equation}
  \omega_{ij} = \frac12k_T^2\frac{p_i\cdot p_j}{(p_i\cdot k)(p_j\cdot k)},
\end{equation}
followed by further evolution from $k_T$ down to $Q_0$.  The real
emission contribution on the other hand involves the matrix
$\mathbf{D}^\mu$, which describes the emission of a gluon with Lorentz
index $\mu$ and is rectangular, being the transformation from the
2-dimensional colour space of $qq\to qq$ to the 4-dimensional colour
space of $qq\to qqg$.  We again work in a basis for the process
$q_iq_j\to q_kq_lg_a$ that differs from the one in \cite{Forshaw:2006fk} only in its
normalization:
\begin{eqnarray}
  \mathbf{C}_1 &=& \frac1{\sqrt{N(N^2-1)}}
  \Bigl(T_{ki}^a\delta_{lj}+T_{lj}^a\delta_{ki}\Bigr), \\
  \mathbf{C}_2 &=& \frac{2\sqrt{N}}{\sqrt{(N^2-1)(N^2-4)}}
  \Bigl(T_{ki}^bT_{lj}^c\,d^{abc}\Bigr), \\
  \mathbf{C}_3 &=& \frac1{\sqrt{N(N^2-1)}}
  \Bigl(T_{ki}^a\delta_{lj}-T_{lj}^a\delta_{ki}\Bigr), \\
  \mathbf{C}_4 &=& \frac2{\sqrt{N(N^2-1)}}
  \Bigl(T_{ki}^bT_{lj}^c\,if^{abc}\Bigr).
\end{eqnarray}
We then have
\begin{equation}
  \mathbf{D}^\mu = \left(\begin{array}{cc}
    \frac{\sqrt{N^2-1}}{2\sqrt{N}}(-h_1^\mu-h_2^\mu+h_3^\mu+h_4^\mu) &
    \frac1{2\sqrt{N}}(-h_1^\mu-h_2^\mu+h_3^\mu+h_4^\mu) \\
    0 &
    \frac{\sqrt{N^2-4}}{2\sqrt{N}}(-h_1^\mu-h_2^\mu+h_3^\mu+h_4^\mu) \\
    \frac{\sqrt{N^2-1}}{2\sqrt{N}}(-h_1^\mu+h_2^\mu+h_3^\mu-h_4^\mu) &
    \frac1{2\sqrt{N}}(h_1^\mu-h_2^\mu-h_3^\mu+h_4^\mu) \\
    0 &
    \frac{\sqrt{N}}2(-h_1^\mu+h_2^\mu-h_3^\mu+h_4^\mu)
  \end{array}\right),
\end{equation}
with
\begin{equation}
  h_i^\mu = \frac12k_T\frac{p_i^\mu}{p_i\cdot k}.
\end{equation}
Note that $\omega_{ij}=2h_i\cdot h_j$ and $\mathbf{D}_\mu^\dagger\mathbf{D}^\mu=-2\boldsymbol{\gamma}$.  
Finally, we need the anomalous dimension matrix for the evolution of this five-parton
system.  This was calculated in \cite{Kyrieleis:2005dt} and, in the
normalized basis, is given by
\begin{eqnarray}
  \mathbf{\Lambda} &=& \left(\begin{array}{cccc}
    \frac{N}{4}(Y-i\pi)+\frac{1}{2N}i\pi &
    \frac{\sqrt{N^2-4}}{2N}i\pi &
    -\frac{N}{4}s_{y}Y &
    0 \\
    \frac{\sqrt{N^2-4}}{2N}i\pi &
    \frac{N}{4}(2Y-i\pi)-\frac{3}{2N}i\pi &
    0 &
    0 \\
    -\frac{N}{4}s_{y}Y &
    0 &
    \frac{N}{4}(Y-i\pi)-\frac{1}{2N}i\pi &
    -\frac{1}{2}i\pi \\
    0 &
    0 &
    -\frac{1}{2}i\pi &
    \frac{N}{4}(2Y-i\pi)-\frac{1}{2N}i\pi
  \end{array}\right) \nonumber\\
  &+& \left(\begin{array}{cccc}
    N & 0 & 0 & 0\\
    0 & N & 0 & 0\\
    0 & 0 & N & 0\\
    0 & 0 & 0 & N
  \end{array}\right)\frac{1}{4}\rho(Y,2|y|)
  + \left(\begin{array}{cccc}
    C_{F} & 0 & 0 & 0\\
    0 & C_{F} & 0 & 0\\
    0 & 0 & C_{F} & 0\\
    0 & 0 & 0 & C_{F}
  \end{array}\right)\frac{1}{2}\rho(Y,\Delta y) \nonumber\\
  &+& \left(\begin{array}{cccc}
    -\frac{N}4 & 0 & -\frac{N}4s_y & \frac12s_y \\
    0 & -\frac{N}4 & 0 & \frac{\sqrt{N^2-4}}4s_y \\
    -\frac{N}4s_y & 0 & -\frac{N}4 & -\frac12 \\
    \frac12s_y & \frac{\sqrt{N^2-4}}4s_y & -\frac12 & -\frac{N}4
  \end{array}\right)\frac12\lambda,
\end{eqnarray}
where $s_y=\mathrm{sgn}(y)$ and, like $\rho$, $\lambda$ is small in the
region of interest (it is defined in \cite{Forshaw:2006fk}).  The only
property that we shall need here is that, in the final-state collinear
limit, $\lambda=\rho(Y,2|y|)=\rho(Y,\Delta y)$.

We are now ready to
calculate $\sigma_1$.  The all-orders calculation was discussed in
\cite{Forshaw:2006fk}.  Here we focus on its order-by-order
expansion in the initial-state collinear limit. Specifically we consider the out-of-gap gluon to be collinear to the incoming quark with momentum $p_1$, i.e. $y\to\infty$ at fixed
$k_T$.  Here $\boldsymbol{\gamma}$ and $\mathbf{D}^\mu$ simplify, since
$h_1\gg h_2\sim h_3\sim h_4\equiv h$, implying corresponding results for
$\omega_{ij}$.  We obtain two contributions, which it will be useful to keep
separate in anticipation of our comparison with the result obtained using the basis
independent method.  The first contribution is when the out-of-gap gluon (either real or virtual)
is hardest. By that we mean that is has the largest $k_T$ of all the soft gluons that dress the primary hard scatter. It is accompanied by two Coulomb gluons and an
eikonal gluon, all at lower $k_T$.  Replacing the $y$ integration by $2\ln (Q/k_T)$, as
explained in \cite{Forshaw:2006fk}, and integrating over $k_T$ we obtain
\begin{equation}
  \sigma_{1,\mbox{\tiny out=hardest}} =
  -\sigma_0\left(\frac{2\alpha_s}\pi\right)^4
  \ln^5\left(\frac{Q}{Q_0}\right)\pi^2Y\frac{N^2-2}{240}\,. \label{eq:reshardest}
\end{equation}
The second contribution comes when the out-of-gap gluon is the second
hardest. It is accompanied by one harder Coulomb gluon and two lower $k_T$ gluons (one eikonal and one Coulomb). In this case we obtain
\begin{equation}
  \sigma_{1,\mbox{\tiny out=second-hardest}} =
  -\sigma_0\left(\frac{2\alpha_s}\pi\right)^4
  \ln^5\left(\frac{Q}{Q_0}\right)\pi^2Y\frac{N^2-1}{120}\,. \label{eq:resnexthardest}
\end{equation}
Summing the two contributions, we obtain the result in equation (3.24) of
\cite{Forshaw:2006fk} (up to a factor of two, since the result
there is for an out-of-gap gluon with $y>0$, while the factor of 2 introduced in the $y$
integration above accounts for the fact that the out-of-gap gluon could
be on either side of the gap):
\begin{equation}
  \sigma_1 =
  -\sigma_0\left(\frac{2\alpha_s}\pi\right)^4
  \ln^5\left(\frac{Q}{Q_0}\right)\pi^2Y\frac{3N^2-4}{240}\,.
\end{equation}
We will use these results in Section \ref{onegluon}, as a cross-check of
the basis independent results.

\section{Colour Basis Independent Notation}
\label{colour}

The colour basis independent notation we use was developed
by Catani, Marchesini and others (for examples, see
\cite{Catani:1985xt,Catani:1996jh,Catani:1996vz,Dokshitzer:2005ig,Dokshitzer:2005ek}).
In this section we introduce it before using it to derive the anomalous dimension matrix for a rapidity gap in
an arbitrary ($m$-parton) final state.

We start by considering the amplitude for the emission of a soft gluon off an
$m$-parton amplitude. We can write
\begin{equation}
|m+1 \rangle = g \sum_i \frac{p_i \cdot \epsilon^*}{p_i \cdot k} \; \mathbf{T}_i^a \, |m \rangle
\label{eq:sge}
\end{equation}
where $g$ is the strong coupling constant, $k$ is the momentum of the emitted
gluon and $\epsilon$ is its polarization vector. The ket $|m\rangle$ represents the
amplitude prior to emission and makes explicit that it is a vector in colour space. The space is spanned by a set of basis kets $|i\rangle$ which we can take to form an orthonormal set. $\mathbf{T}_i^a$ is the operator that determines the map from the $m$ dimensional vector space onto the $m+1$ dimensional space which occurs as a result of emitting a gluon of colour $a$. One might choose to represent the $m$-parton amplitude by $M_{i_1 i_2 i_3\cdots i_m}$ where the indices are the colour indices of incoming or outgoing quarks, antiquarks or gluons. In such a representation, the $\mathbf{T}_i^a$ in Eq.(\ref{eq:sge}) will be represented by $\mathbf{t}^a$, the generator in the fundamental representation, if parton $i$ is either an outgoing quark or an incoming antiquark. The sign reverses if $i$ is an incoming quark or an outgoing antiquark. Similarly, if the radiating parton is a gluon we should use the generator in the adjoint representation, $-i \mathbf{f}^a$. In all cases, it is to be understood that Eq.(\ref{eq:sge}) provides the definition of the sign convention of the soft gluon emission vertex. The Hermitian conjugate operator $(\mathbf{T}_i^a)^{\dagger}$ determines the map from the $m+1$ dimensional vector space to the $m$ dimensional space corresponding to the absorption of a gluon of colour $a$.  

Under a general $SU(3)$ transformation, $|m\rangle$ transforms as a colour singlet and since the generators in the $m$-parton representation correspond to $\sum_{i=1}^{m} \mathbf{T}_i^a$ it follows that
\begin{equation}
 \sum_{i=1}^m \mathbf{T}_i^a | m \rangle = 0.
 \end{equation}
This identity will prove to be very useful. Also of note is the fact that
\begin{equation}
(\mathbf{T}_i^a)^{\dagger} \mathbf{T}_j^a = (\mathbf{T}_j^a)^{\dagger} \mathbf{T}_i^a,
\end{equation}
as a result of which we introduce the notation
\begin{equation}
(\mathbf{T}_i^a)^{\dagger} \mathbf{T}_j^a \equiv \mathbf{T}_i \cdot \mathbf{T}_j~.
\end{equation}

In this basis independent framework, we can write down the anomalous dimension matrix for an arbitrary phase-space veto in an arbitrary $m$-parton process. It is
\begin{equation}
  \label{generalGamma}
  \mathbf{\Gamma} = -\sum_{i<j} \mathbf{T}_i\cdot\mathbf{T}_j\;
  \Omega_{ij},
\end{equation}
where
\begin{equation}
  \Omega_{ij} = \frac12\left\{
    \int_{\mbox{\tiny veto}} \frac{dy\,d\phi}{2\pi}\omega_{ij}
    -i\pi\,\Theta(ij=II\mbox{~or~}FF)
  \right\},
\end{equation}
the sum over $i,j$ runs over all partons in the initial and final state
and the ordering $i<j$ is simply to ensure that each pair is counted
once, since $\mathbf{T}_i\cdot\mathbf{T}_j$ and $\Omega_{ij}$ are both
symmetric under interchange of $i$ and $j$.\footnote{We work in Feynman gauge and assume massless partons, thereby avoiding self-energy contributions with $i=j$.}  
Note that, apart from the sign convention ($\Omega_{ij}$ always takes a plus sign in this paper)
this is the same definition as that used in \cite{Kyrieleis:2005dt}. The theta function
ensures that the $i \pi$ contribution is present only when $ij$ correspond
to a pair of incoming ($II$) or outgoing ($FF$) partons.

$\Omega_{ij}$ was calculated in \cite{Kyrieleis:2005dt} for the
case of an azimuthally-symmetric gap in rapidity of length $Y$, with
sufficient generality for arbitrary $i,j$ kinematics.  It can be summarized as
\begin{eqnarray}
  \Omega_{ij} &=& \frac12\left\{
    Y\,\Theta(\mbox{$ij$~on~opposite~sides~of~gap})
    +\frac12\rho(Y;2|y_i|)+\frac12\rho(Y;2|y_j|)
\right.\nonumber\\&&\left.\phantom{\frac12}
    -\lambda(Y;|y_i|+|y_j|,|\phi_i-\phi_j|)
    \Theta(\mbox{$ij$~on~same~side~of~gap})
    -i\pi\,\Theta(ij=II\mbox{~or~}FF)
  \right\},
\nonumber\\&&
\end{eqnarray}
where $\rho$ and $\lambda$ are known functions, the only properties of
which we will need here are:
\begin{enumerate}
\item $\rho(Y;|y|),\lambda(Y;|y|,\Delta\phi)\to0$ as $|y|\to\infty$; and
\item $\lambda(Y;|y|,\Delta\phi)\to\rho(Y;|y|)$ as $\Delta\phi\to0$.
\end{enumerate}
The first property ensures that $\rho$ and $\lambda$ are absent for
initial-state partons.

The simplicity of the result for $\Omega_{ij}$ allows one to simplify the
result for $\mathbf{\Gamma}$.  Gathering together terms
with the same momentum dependence, we obtain
\begin{eqnarray}
\hspace*{-1em}
  \mathbf{\Gamma} &=&
  -\frac12Y\left(\sum_{i\in L}\mathbf{T}_i\right)\cdot
  \left(\sum_{j\in R}\mathbf{T}_j\right)
  +\frac12i\pi\left(\mathbf{T}_1\cdot\mathbf{T}_2
  +\sum_{(i<j)\in F}\mathbf{T}_i\cdot\mathbf{T}_j\right)
\nonumber\\&&
  -\frac14\sum_{i\in F}\rho(Y;2|y_i|)
  \sum_{j\not=i}\mathbf{T}_i\cdot\mathbf{T}_j
\nonumber\\&&
  +\frac12\sum_{(i<j)\in L}\lambda(Y;|y_i|+|y_j|,|\phi_i-\phi_j|)
  \mathbf{T}_i\cdot\mathbf{T}_j
  +\frac12\sum_{(i<j)\in R}\lambda(Y;|y_i|+|y_j|,|\phi_i-\phi_j|)
  \mathbf{T}_i\cdot\mathbf{T}_j,
\hspace*{-1em}
\nonumber\\[-2ex]
\end{eqnarray}
where the labels $L$ and $R$ label the bunches of partons on each side 
of the gap and the indices ``1" and ``2" refer to the two incoming partons.  
Now we can use colour conservation to simplify
this expression further.  Firstly we have
\begin{equation}
  \sum_{j\not=i}\mathbf{T}_i\cdot\mathbf{T}_j=-\mathbf{T}_i^2,
\end{equation}
and secondly we can perform a similar trick on the Coulomb
gluon terms:
\begin{equation}
  \sum_{(i<j)\in F}\mathbf{T}_i\cdot\mathbf{T}_j
  =\mathbf{T}_1\cdot\mathbf{T}_2
  +\frac12\left(\sum_{i\in I}\mathbf{T}_i^2
  -\sum_{i\in F}\mathbf{T}_i^2\right). \label{eq:id1}
\end{equation}
Now, adding an imaginary multiple of the identity matrix to the anomalous dimension
matrix has no physical effect, so we are free to drop the
$\mathbf{T}_i^2$ terms.  The Coulomb gluon terms
are thus proportional to $\mathbf{T}_1\cdot\mathbf{T}_2$ and are
absent if one or both of the incoming partons is colourless.  This was
first pointed out using the colour basis independent notation in \cite{Catani:1985xt} and is an important component of the proofs of factorization in \cite{Collins:1985ue,Collins88,Collins:1998ps}.

Finally, we can re-write the leading ($\sim Y$) eikonal gluon term using
\begin{equation}
  \left(\sum_{i\in L}\mathbf{T}_i\right)\cdot
  \left(\sum_{j\in R}\mathbf{T}_j\right)
  =-\left(\sum_{i\in L}\mathbf{T}_i\right)^2
  =-\left(\sum_{i\in R}\mathbf{T}_i\right)^2
  \equiv-\mathbf{T}_t^2.
\end{equation}
That is, if we think of the rapidity gap as separating the partonic
event into two separate systems, the dominant Sudakov suppression
effectively comes from emission off the total colour charge exchanged
between the two systems, as noticed in \cite{Forshaw:2007vb}.

Thus the full result for the anomalous dimension matrix for an
azimuthally symmetric rapidity gap of length $Y$ is given by
\begin{eqnarray}
\hspace*{-1em}
  \mathbf{\Gamma} &=&
  \frac12Y\mathbf{T}_t^2
  +i\pi\mathbf{T}_1\cdot\mathbf{T}_2
  +\frac14\sum_{i\in F}\rho(Y;2|y_i|)\mathbf{T}_i^2
\nonumber\\&&
  +\frac12\sum_{(i<j)\in L}\lambda(Y;|y_i|+|y_j|,|\phi_i-\phi_j|)
  \mathbf{T}_i\cdot\mathbf{T}_j
  +\frac12\sum_{(i<j)\in R}\lambda(Y;|y_i|+|y_j|,|\phi_i-\phi_j|)
  \mathbf{T}_i\cdot\mathbf{T}_j ~.
\hspace*{-1em}
\nonumber\\[-2ex] \label{eq:gamma}
\end{eqnarray}
Notice that the terms involving $\rho$ are Abelian in nature since $\mathbf{T}_i^2 = C_F \boldsymbol{\mathbbm{1}}$ or $C_A \boldsymbol{\mathbbm{1}}$ depending upon whether parton $i$ is a quark/antiquark or gluon.

\begin{figure}[t]
\centering
\subfigure[]{\includegraphics[width=0.63\textwidth]{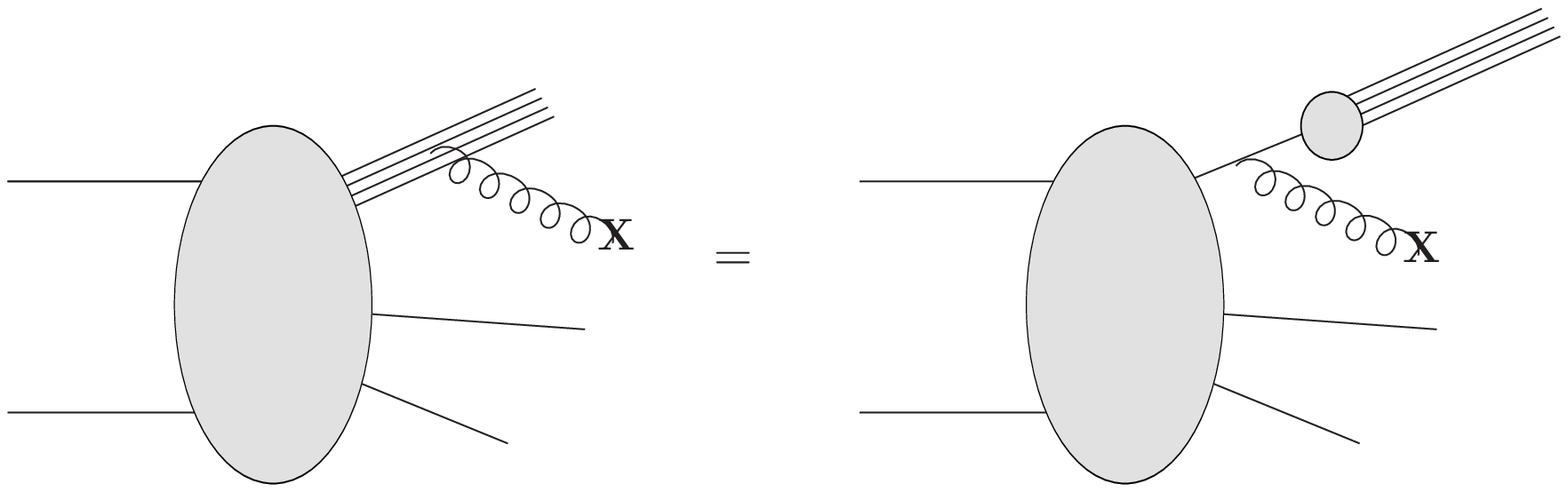}}\\
\subfigure[]{\includegraphics[width=0.9\textwidth]{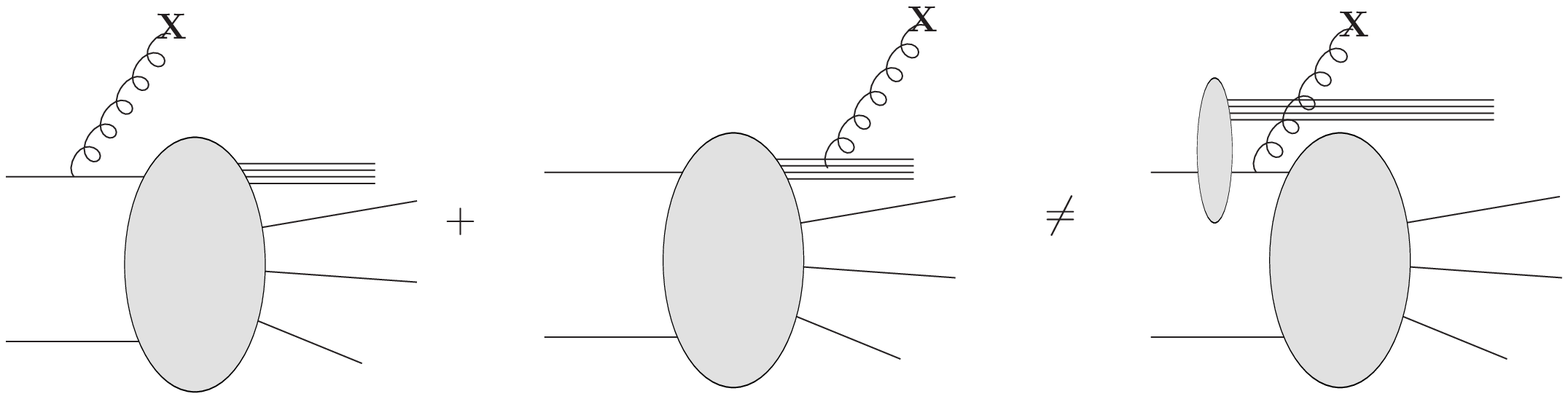}} 
\caption{Factorization of soft gluon emission off a collinear bunch of
partons.  The cross indicates that the gluon can be attached to any other external leg. A sum over couplings to the final state collinear partons is implied.}
\label{fig:factorize}
\end{figure}

Now we can prove a very important property of $\mathbf{\Gamma}$: it
is safe against final state collinear singularities.  More specifically, if any
two or more partons in the final state become collinear with each
other, the soft gluon evolution of the system is identical to the evolution 
of the system in which the collinear partons are replaced by a single parton with the
same total colour charge.  That is, if $k$ and $l$ are the collinear partons, then
$\mathbf{\Gamma}$ depends only upon $\mathbf{T}_k+\mathbf{T}_l$ and not upon the
$\mathbf{T}_k$ or $\mathbf{T}_l$ separately. 
The proof is straightforward. Let us consider partons $k$ and $l$ to be final state and collinear\footnote{The generalization to more than two collinear partons is straightforward.}.
We first note that since the imaginary part of $\mathbf{\Gamma}$ can be
written in terms of $\mathbf{T}_1\cdot\mathbf{T}_2$ only, it has no
explicit dependence on $\mathbf{T}_k$ and $\mathbf{T}_l$ and we need
only consider the real part of $\mathbf{\Gamma}$.
The part of  $\mathbf{\Gamma}$ that depends upon
the colour charges of $k$ and $l$ is
\begin{equation}
 \mathrm{I\!Re}(\mathbf{\Gamma}^{(kl)}) = \sum_{i \neq k,l}  \mathbf{T}_i\cdot\mathbf{T}_k \; \mathrm{I\!Re}(\Omega_{ik})
 + \sum_{i \neq k,l} \mathbf{T}_i\cdot\mathbf{T}_l \; \mathrm{I\!Re}(\Omega_{il})
 + \mathbf{T}_k\cdot\mathbf{T}_l \; \mathrm{I\!Re}(\Omega_{kl}).
 \end{equation}
Now since $k$ and $l$ are collinear $\mathrm{I\!Re}(\Omega_{ik}) = \mathrm{I\!Re}(\Omega_{il})$. The equality follows since $\omega_{ij}$ depends only on the direction of partons $i$ and $j$ and not their energies.
Moreover, $\mathrm{I\!Re}(\Omega_{kl})$ vanishes in the collinear limit
since the numerator of $\omega_{kl}$ vanishes.
It now follows immediately that $ \mathbf{\Gamma}^{(kl)}$ depends only upon the sum $\mathbf{T}_k+\mathbf{T}_l$ and hence that soft gluons factorize from collinear final state emissions as illustrated in Figure \ref{fig:factorize}(a).

We contrast this result with that in the initial state collinear limit,
in which one or more outgoing partons becomes collinear with one of the
incoming partons.  Precisely the same reasoning as before can be used for the real
part of $\mathbf{\Gamma}$, but since the imaginary part can be written
in terms of the colours of the initial state partons only, it does not
depend on the sum of the colour charges of the collinear partons and the
factorization is broken, as illustrated in Figure \ref{fig:factorize}(b).

It is this fact that we described in
\cite{Kyrieleis:2006fh,Seymour:2007vw} as a breakdown of naive
coherence.  It leads directly to the appearance of super-leading
logarithms in the calculation of the gaps-between-jets cross-section.

\section{One Gluon Outside the Gap}
\label{onegluon}

We now proceed to calculate the one gluon outside the gap cross-section
according to Eqs.~(\ref{master})--(\ref{masterR}) in the colour basis
independent notation.  Since we require the colour evolution of the $m$
and $m\!+\!1$ parton systems, we modify the notation slightly, to
differentiate between the colour matrix to emit a gluon from a parton
$i$ in the $m$ parton system, $\mathbf{t}_i$, and in the $m\!+\!1$
parton system, $\mathbf{T}_i$.  We continue to use the notation from
\cite{Forshaw:2006fk} and \cite{Kyrieleis:2005dt} in which the
additional gluon in the $m\!+\!1$-parton system is labelled $k$, whilst
the others are labelled by indices running over the range 1 to~$m$.

Without loss of generality, we assume that the gluon is emitted on the
same side of the gap as partons 1 and~3 and, from Eq.(\ref{eq:gamma}), it follows that
\begin{eqnarray}
  \mathbf{\Gamma} &=&
  \frac12Y\mathbf{t}_t^2
  +i\pi\mathbf{t}_1\cdot\mathbf{t}_2
  +\frac14\rho(Y;\Delta y)\left(\mathbf{t}_3^2+\mathbf{t}_4^2\right), \\
  \mathbf{\Lambda} &=&
  \frac12Y\mathbf{T}_t^2
  +i\pi\mathbf{T}_1\cdot\mathbf{T}_2
  +\frac14\rho(Y;\Delta y)\left(\mathbf{T}_3^2+\mathbf{T}_4^2\right)
  +\frac14\rho(Y;2y)\mathbf{T}_k^2
\nonumber\\&&
  +\frac12\lambda(Y;\mbox{$\frac12$}\Delta y+y,\phi)
  \mathbf{T}_3\cdot\mathbf{T}_k,
\end{eqnarray}
with $\mathbf{t}_t^2=(\mathbf{t}_1+\mathbf{t}_3)^2=
(\mathbf{t}_2+\mathbf{t}_4)^2$ and
$\mathbf{T}_t^2=(\mathbf{T}_1+\mathbf{T}_3+\mathbf{T}_k)^2=
(\mathbf{T}_2+\mathbf{T}_4)^2$.
We also require the real and virtual emission matrices, which are given by
\begin{eqnarray}
  \mathbf{D}^\mu_a &=& \sum_i\mathbf{t}_i^a\;h_i^\mu, \\
  \boldsymbol{\gamma} &=&
  -\frac12\sum_{i<j}\mathbf{t}_i\cdot\mathbf{t}_j\;\omega_{ij}.
\end{eqnarray}
Now we specialize to the case where the gluon is collinear with incoming parton~1, i.e. it has $y\to\infty$.
In this case, the last two terms of $\mathbf{\Lambda}$ vanish.  The remaining
$\rho$ terms are once again Abelian, and the same for both final states, so they lead only to an overall factor, which we neglect in the following discussion.  The real and virtual emission matrices also
simplify in this limit, and we have
\begin{eqnarray}
  \mathbf{\Gamma} &=&
  \frac12Y\mathbf{t}_t^2
  +i\pi\mathbf{t}_1\cdot\mathbf{t}_2, \\
  \mathbf{\Lambda} &=&
  \frac12Y\mathbf{T}_t^2
  +i\pi\mathbf{T}_1\cdot\mathbf{T}_2, \\
  \mathbf{D}^\mu_a &=& (h_1^\mu-h^\mu)\mathbf{t}_1^a, \label{hmu}\\
  \boldsymbol{\gamma} &=& \frac{1}{2} \mathbf{t}_1^2.
\end{eqnarray}
In Eq.~(\ref{hmu}) we have introduced $h^\mu$ to emphasize the fact that
in the collinear limit all the $h_i^\mu$ for $i\not=1$ are equal.
Note that we have used $(h_1-h)\cdot (h_1-h) = -1$ to simplify $\boldsymbol{\gamma}$.
The cross-section can then be written as
\begin{eqnarray}
  \label{fullresult}
  \sigma_1 = -\frac{2\alpha_s}{\pi}\int_{Q_0}^Q\frac{dk_T}{k_T}
  \left(2\ln\frac{Q}{k_T}\right)
  \Biggl\langle\!m_0\Biggr|
  &&e^{-\frac{2\alpha_s}{\pi}\int_{k_T}^Q\frac{dk_T'}{k_T'}
    \left(\frac12Y\mathbf{t}_t^2
    -i\pi\mathbf{t}_1\cdot\mathbf{t}_2\right)}
\nonumber\\&&\hspace*{-3.8cm}
  \Biggl\{
  \mathbf{t}_1^2
  e^{-\frac{2\alpha_s}{\pi}\int_{Q_0}^{k_T}\frac{dk_T'}{k_T'}
    \left(\frac12Y\mathbf{t}_t^2
    -i\pi\mathbf{t}_1\cdot\mathbf{t}_2\right)}
  e^{-\frac{2\alpha_s}{\pi}\int_{Q_0}^{k_T}\frac{dk_T'}{k_T'}
    \left(\frac12Y\mathbf{t}_t^2
    +i\pi\mathbf{t}_1\cdot\mathbf{t}_2\right)}
\nonumber\\&&\hspace*{-3.2cm}
  -\mathbf{t}_1^{a \dagger}
  e^{-\frac{2\alpha_s}{\pi}\int_{Q_0}^{k_T}\frac{dk_T'}{k_T'}
    \left(\frac12Y\mathbf{T}_t^2
  -i\pi\mathbf{T}_1\cdot\mathbf{T}_2\right)}
  e^{-\frac{2\alpha_s}{\pi}\int_{Q_0}^{k_T}\frac{dk_T'}{k_T'}
    \left(\frac12Y\mathbf{T}_t^2
  +i\pi\mathbf{T}_1\cdot\mathbf{T}_2\right)}
  \mathbf{t}_1^a
  \Biggr\}
\nonumber\\&&
  e^{-\frac{2\alpha_s}{\pi}\int_{k_T}^Q\frac{dk_T'}{k_T'}
    \left(\frac12Y\mathbf{t}_t^2
    +i\pi\mathbf{t}_1\cdot\mathbf{t}_2\right)}
  \Biggl|m_0\!\Biggr\rangle.
\end{eqnarray}
Note that it is the non-commutativity of $\mathbf{T}_t^2$ and
$\mathbf{T}_1\cdot\mathbf{T}_2$ (and similarly $\mathbf{t}_t^2$ and
$\mathbf{t}_1\cdot\mathbf{t}_2$) that prevents this expression from
cancelling to zero: if they commuted then the two exponentials could be
combined, all $\mathbf{T}_1\cdot\mathbf{T}_2$ and
$\mathbf{t}_1\cdot\mathbf{t}_2$ dependence would cancel, $\mathbf{t}_1$
could be commuted through $\mathbf{T}_t^2$ and the real and virtual
parts would be identical.

To find the first non-zero super-leading logarithm, we expand the main
bracket of this expression order by order in $\alpha_s$:
\begin{equation}
  \Biggl\{\phantom{...}\Biggr\}_0=
  \mathbf{t}_1^2-\mathbf{t}_1^{a \dagger}\mathbf{t}_1^a = 0.
\end{equation}
\begin{equation}
  \Biggl\{\phantom{...}\Biggr\}_1 =
  -\frac{2\alpha_s}{\pi}\int_{Q_0}^{k_T}\frac{dk_T'}{k_T'}
  \Biggl\{\mathbf{t}_1^2Y\mathbf{t}_t^2
  -\mathbf{t}_1^{a \dagger} Y\mathbf{T}_t^2\,\mathbf{t}_1^a\Biggr\}
\end{equation}
is also zero because 
\begin{equation}
  \mathbf{T}_t^2\,\mathbf{t}_1^a=\mathbf{t}_1^a\,\mathbf{t}_t^2.
\end{equation}
Expanding to order $\alpha_s^2$ yields
\begin{eqnarray}
  \Biggl\{\phantom{...}\Biggr\}_2 &=&
  \left( \frac{i\pi Y}{2}\right)
  \left(-\frac{2\alpha_s}{\pi}\int_{Q_0}^{k_T}\frac{dk_T'}{k_T'}\right)^2
  \Biggl\{
  \mathbf{t}_1^2
    \left[\mathbf{t}_t^2\, , \mathbf{t}_1\cdot\mathbf{t}_2 \right]
  -\mathbf{t}_1^{a\dagger}
    \left[\mathbf{T}_t^2\,, \mathbf{T}_1\cdot\mathbf{T}_2 \right]
  \mathbf{t}_1^a \label{eq:a2}
  \Biggr\} .  \nonumber \\ &&
\end{eqnarray}
Note that this result comes only from the case where there is one Coulomb gluon and one eikonal gluon 
either side of the cut. Now, this term is not zero but the corresponding matrix element is zero, i.e.
\begin{equation}
\langle m_0 | \left\{ ~ \right\}_2 | m_0 \rangle = 0.
\end{equation}
This term will however be relevant at the next order (when we add an additional Coulomb gluon) and so we take the opportunity here to simplify it further.  Using colour conservation, and the fact that
$\mathbf{t}_1\cdot\mathbf{t}_2$ commutes with itself, with all colour
dot products involving only final state particles and with all
$\mathbf{t}_i^2$, we obtain
\begin{equation}
  \left[\mathbf{t}_t^2,\mathbf{t}_1\cdot\mathbf{t}_2\right] = -2
  \left[\mathbf{t}_1\cdot\mathbf{t}_4,\mathbf{t}_1\cdot\mathbf{t}_2\right]
  \quad\mbox{and}\quad
  \left[\mathbf{T}_t^2,\mathbf{T}_1\cdot\mathbf{T}_2\right] = -2
  \left[\mathbf{T}_1\cdot\mathbf{T}_4,\mathbf{T}_1\cdot\mathbf{T}_2\right].
\end{equation}
Thus we can simplify Eq.(\ref{eq:a2}) by replacing the commutator $[\mathbf{t}_t^2,\mathbf{t}_1\cdot\mathbf{t}_2]$ with 
$-2  [\mathbf{t}_1\cdot\mathbf{t}_4,\mathbf{t}_1\cdot\mathbf{t}_2]$, and similarly for $\left[\mathbf{T}_t^2,\mathbf{T}_1\cdot\mathbf{T}_2\right]$.

At the next order we obtain
\begin{eqnarray}
\hspace*{-1em}
  \Biggl\{\phantom{...}\Biggr\}_3 &=&
  \frac1{3!}
  \left(-\frac{2\alpha_s}{\pi}\int_{Q_0}^{k_T}\frac{dk_T'}{k_T'}\right)^3
  \Biggl\{
  \mathbf{t}_1^2
  \biggl(
    \left(Y\mathbf{t}_t^2\right)^3
    +\frac32Y^2i\pi\left(
      \left(\mathbf{t}_t^2\right)^2\mathbf{t}_1\cdot\mathbf{t}_2
     -\mathbf{t}_1\cdot\mathbf{t}_2\left(\mathbf{t}_t^2\right)^2\right)
\nonumber\\&&\hspace*{10em}
    -Y\pi^2\left(
      \mathbf{t}_t^2\,\mathbf{t}_1\cdot\mathbf{t}_2\,
        \mathbf{t}_1\cdot\mathbf{t}_2
      -2\,\mathbf{t}_1\cdot\mathbf{t}_2\,\mathbf{t}_t^2\,
        \mathbf{t}_1\cdot\mathbf{t}_2
      +\mathbf{t}_1\cdot\mathbf{t}_2\,\mathbf{t}_1\cdot\mathbf{t}_2\,
        \mathbf{t}_t^2
    \right)
  \biggr)
\hspace*{-1em}
\nonumber\\&&
  -\mathbf{t}_1^{a \dagger}
  \biggl(
    \left(Y\mathbf{T}_t^2\right)^3
    +\frac32Y^2i\pi\left(
      \left(\mathbf{T}_t^2\right)^2\mathbf{T}_1\cdot\mathbf{T}_2
     -\mathbf{T}_1\cdot\mathbf{T}_2\left(\mathbf{T}_t^2\right)^2\right)
\nonumber\\&&\hspace*{2.5em}
    -Y\pi^2\left(
      \mathbf{T}_t^2\,\mathbf{T}_1\cdot\mathbf{T}_2\,
        \mathbf{T}_1\cdot\mathbf{T}_2
      -2\,\mathbf{T}_1\cdot\mathbf{T}_2\,\mathbf{T}_t^2\,
        \mathbf{T}_1\cdot\mathbf{T}_2
      +\mathbf{T}_1\cdot\mathbf{T}_2\,\mathbf{T}_1\cdot\mathbf{T}_2\,
        \mathbf{T}_t^2
    \right)
  \biggr)
  \mathbf{t}_1^a
  \Biggr\}.
\hspace*{-1em}
\nonumber\\
\end{eqnarray}
Note that the $3!$ seems to imply that only contributions where all three gluons are on the same side of the cut are relevant. This is not the case: the $2Y \pi^2 \, \mathbf{t}_1\cdot\mathbf{t}_2\,\mathbf{t}_t^2 \, \mathbf{t}_1\cdot\mathbf{t}_2$ term actually also has a contribution from when the two Coulomb gluons lie on opposite sides of the cut, i.e. $2/3! = 1/2!-1/3!$. Of the various terms, the first term ($\sim Y^3$) cancels between the real and virtual contributions and the second ($\sim Y^2\pi$ term) vanishes upon forming the matrix element. We are thus left with a leading contribution only from the third ($\sim Y \pi^2$) term:
\begin{eqnarray}
\!
  \Biggl\{\phantom{...}\Biggr\}_3 &\equiv&
  -\frac{Y\pi^2}{6}
  \left(-\frac{2\alpha_s}{\pi}\int_{Q_0}^{k_T}\frac{dk_T'}{k_T'}\right)^3
  \Biggl\{
  \mathbf{t}_1^2
   \Bigl[ \left[
    \mathbf{t}_t^2\,,\mathbf{t}_1\cdot\mathbf{t}_2\right] ,
      \mathbf{t}_1\cdot\mathbf{t}_2 \Bigr] 
\!\nonumber\\&&
  -\mathbf{t}_1^{a \dagger}
  \Bigl[ \left[
    \mathbf{T}_t^2\,, \mathbf{T}_1\cdot\mathbf{T}_2\, \right]  ,
      \mathbf{T}_1\cdot\mathbf{T}_2 \Bigr]
     \mathbf{t}_1^a
  \Biggr\}.
\end{eqnarray}
As before,  $\mathbf{t}_t^2$ can be replaced
by~$-2\,\mathbf{t}_1\cdot\mathbf{t}_4$ (and similarly for $\mathbf{T}_t^2$).

Substituting back into Eq.~(\ref{fullresult}), we obtain a
contribution to the first super-leading logarithm from configurations in
which the out-of-gap gluon is hardest\footnote{Obtained by setting the exponentials that lie outside of the main bracket in Eq.~(\ref{fullresult}) to unity.} of
\begin{eqnarray}
  \sigma_{1,\mbox{\tiny out=hardest}} &=& \left( \frac{2\alpha_s}{\pi} \right)^4 \int_{Q_0}^Q\frac{dk_T}{k_T}
  \left(2\ln\frac{Q}{k_T}\right)  \left( \int_{Q_0}^{k_T}\frac{dk_T'}{k_T'}\right)^3  \; \frac{Y\pi^2}{3}
\nonumber\\&& \hspace*{-2cm}
  \Biggl\langle\!m_0\Biggr|
  \mathbf{t}_1^2
 \Bigl[ \left[
    \mathbf{t}_1 \cdot \mathbf{t}_4 \,,\mathbf{t}_1\cdot\mathbf{t}_2\right] ,
      \mathbf{t}_1\cdot\mathbf{t}_2 \Bigr] 
  -\mathbf{t}_1^{a \dagger}
 \Bigl[ \left[
    \mathbf{T}_1 \cdot \mathbf{T}_4\,, \mathbf{T}_1\cdot\mathbf{T}_2\, \right]  ,
      \mathbf{T}_1\cdot\mathbf{T}_2 \Bigr]
  \mathbf{t}_1^a
  \Biggl|m_0\!\Biggr\rangle. \label{eq:hardest}
\end{eqnarray}
The colour matrix element can be evaluated explicitly and after performing the transverse momentum integrals one  finds the expression for $\sigma_{1,\mbox{\tiny out=hardest}}$ presented in Eq.(\ref{eq:reshardest}).

\begin{figure}[t]
\centering
\subfigure[]{\includegraphics[width=0.4\textwidth]{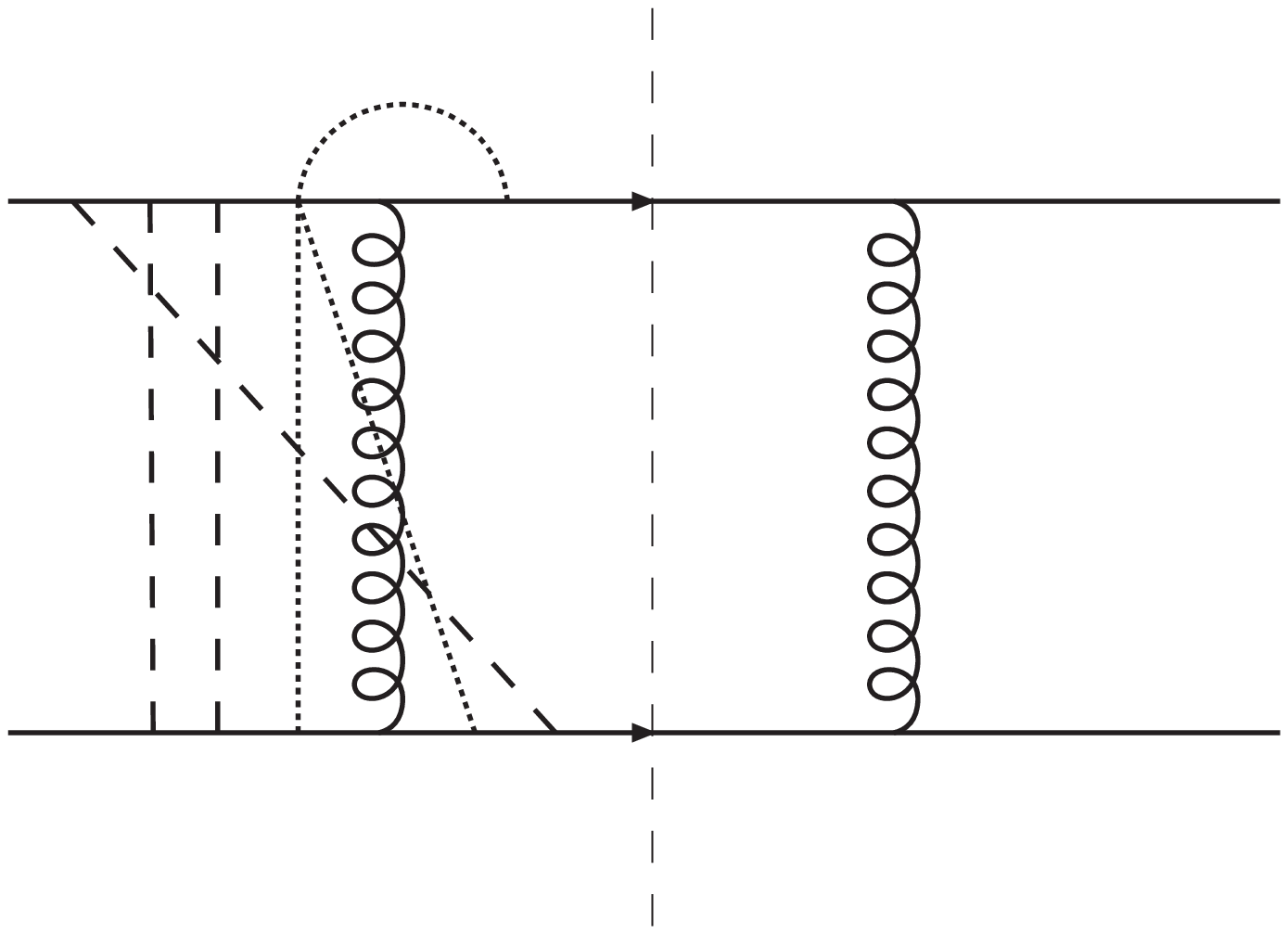}}
\subfigure[]{\includegraphics[width=0.4\textwidth]{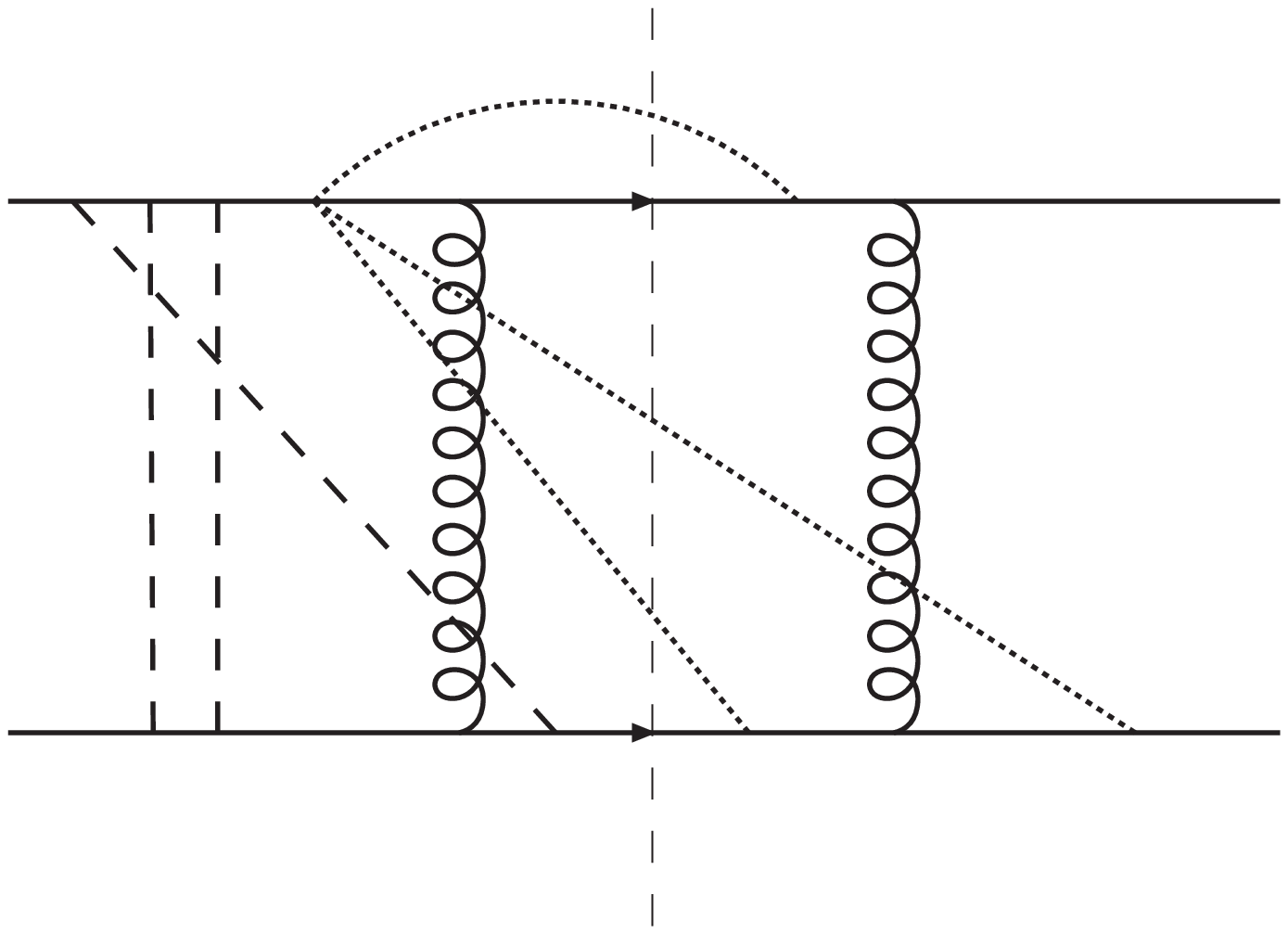}} \\
\subfigure[]{\includegraphics[width=0.4\textwidth]{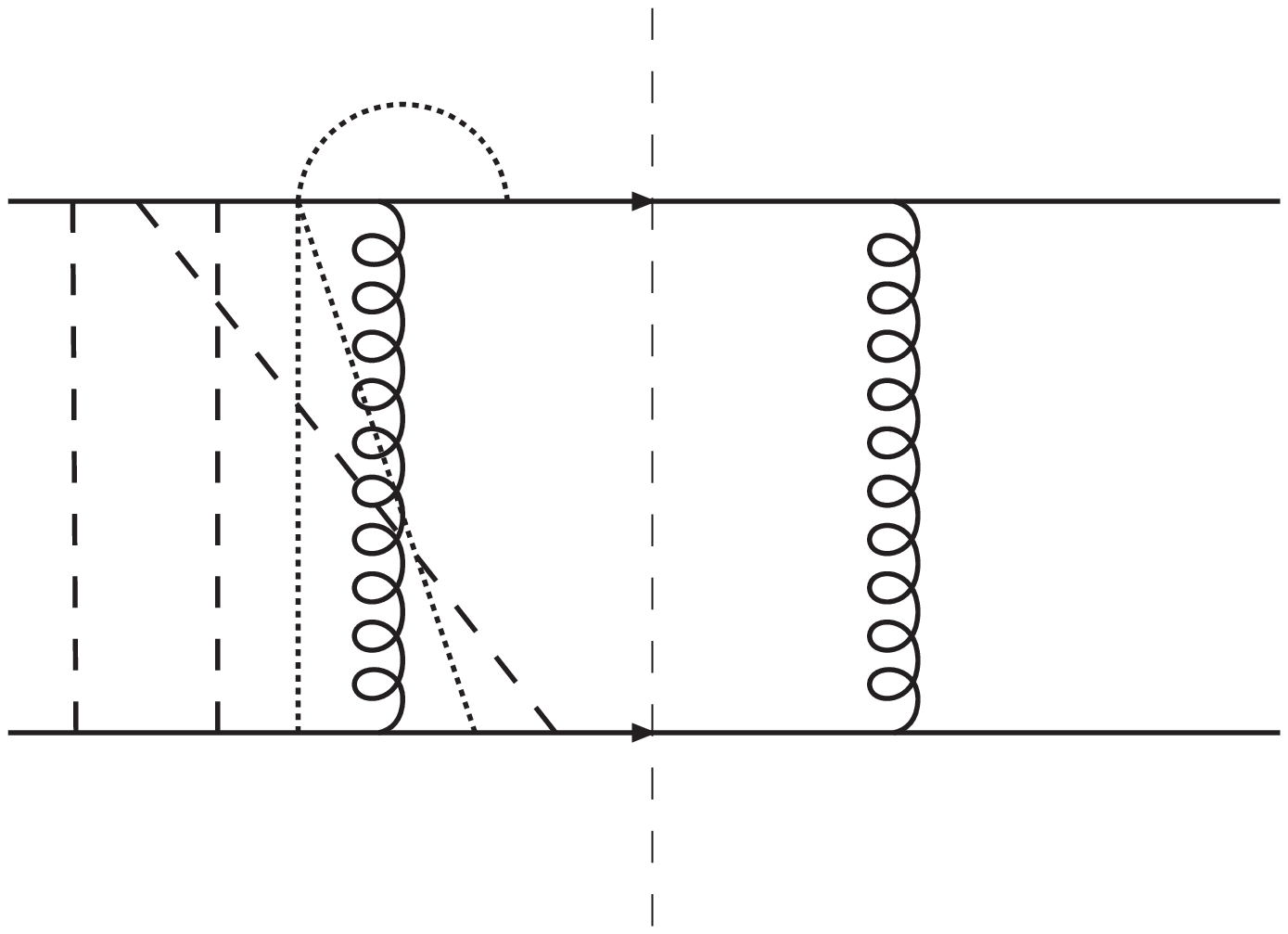}}
\subfigure[]{\includegraphics[width=0.4\textwidth]{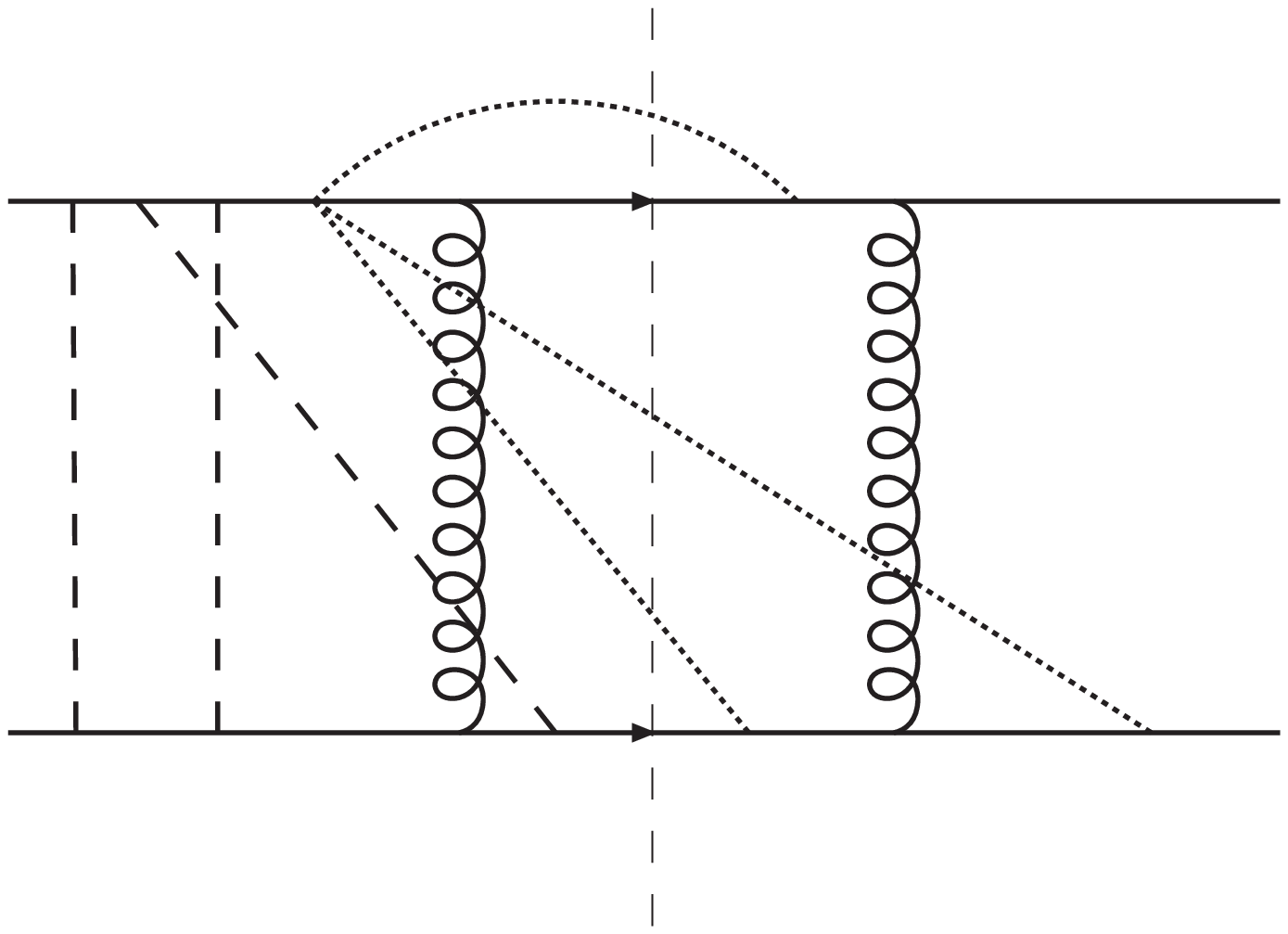}} \\
\subfigure[]{\includegraphics[width=0.4\textwidth]{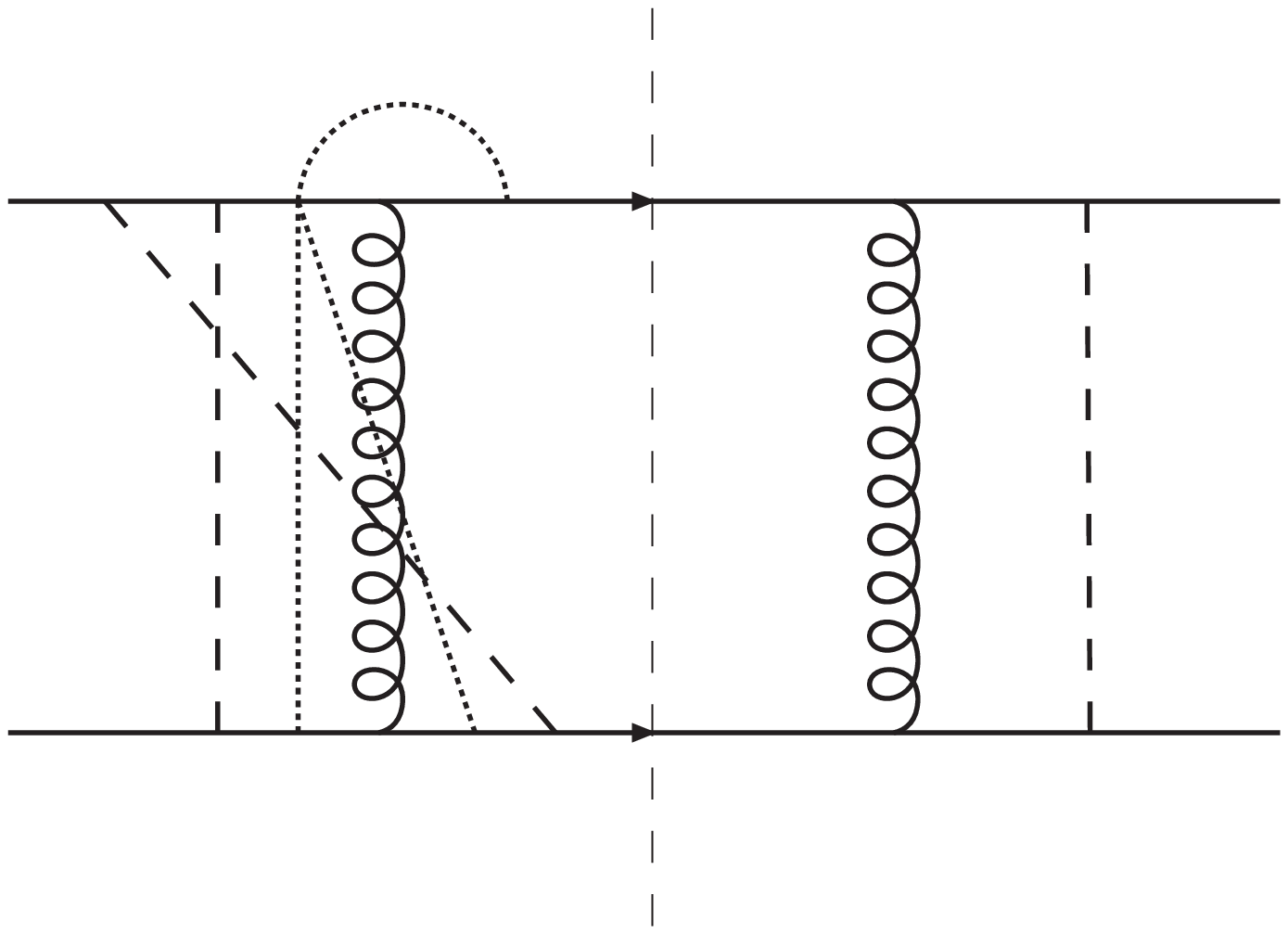}}
\subfigure[]{\includegraphics[width=0.4\textwidth]{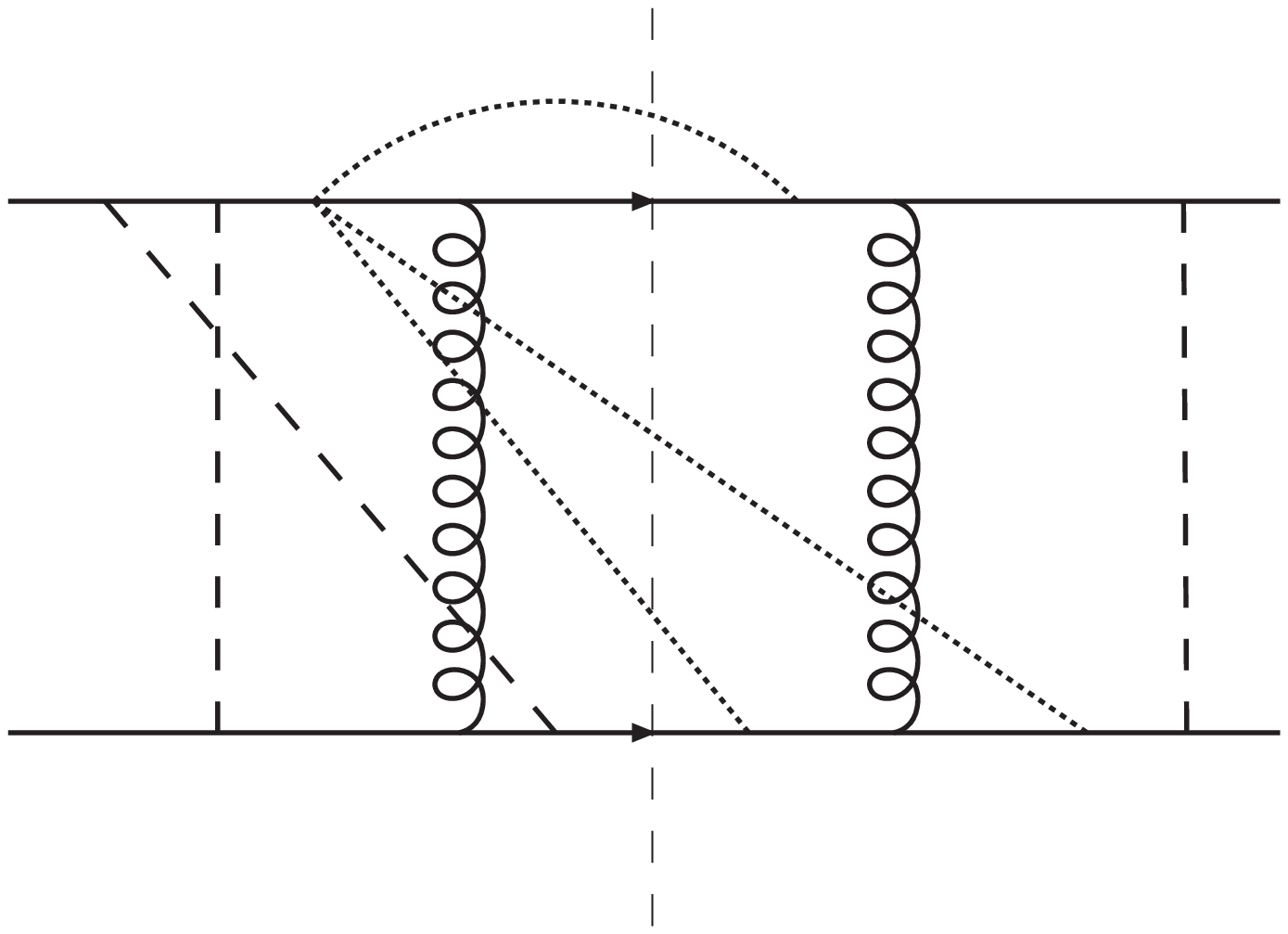}} \\
\subfigure[]{\includegraphics[width=0.4\textwidth]{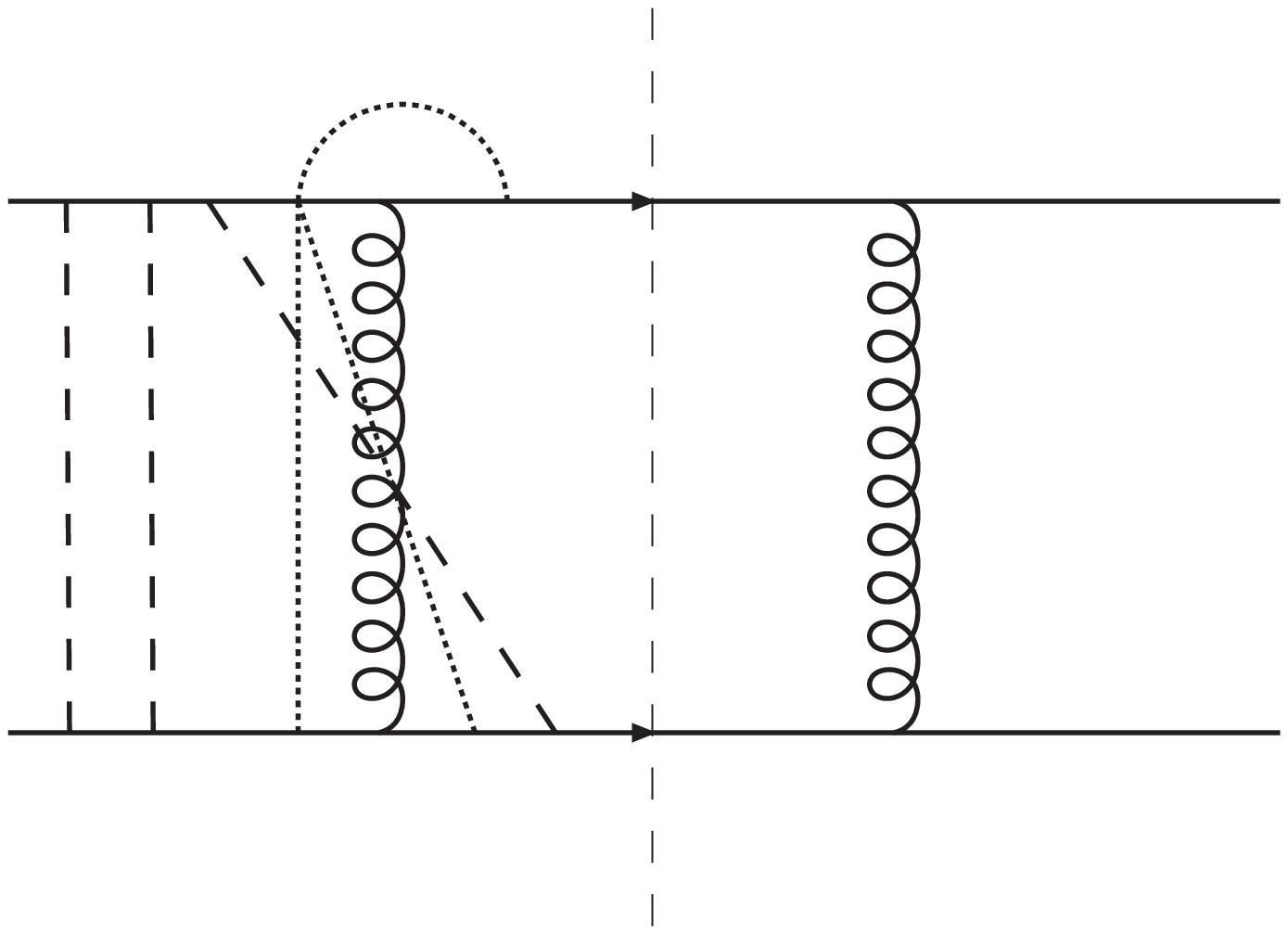}}
\subfigure[]{\includegraphics[width=0.4\textwidth]{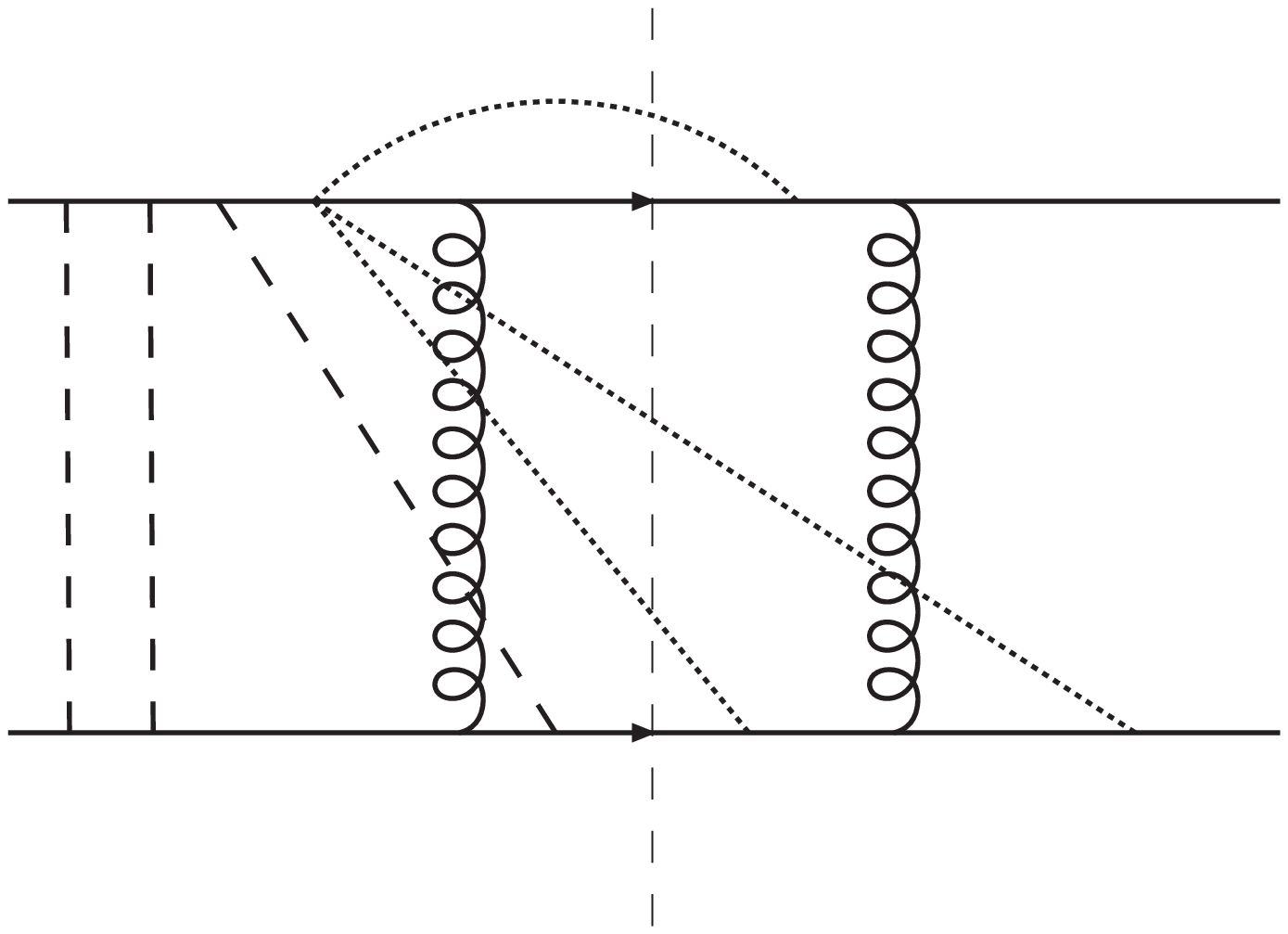}} \\
\caption{The relevant Feynman diagrams in the case that the out-of-gap (dotted) gluon is the hardest gluon. The dashed lines indicate soft (eikonal and Coulomb) gluons. Each subfigure represents three Feynman diagrams, corresponding to the three different ways of attaching the
out-of-gap gluon. In diagrams~(e) and~(f) the soft gluon to the right of the
cut should only be integrated over the region in which it has transverse
momentum less than the out-of-gap gluon.}
\label{fig:hardest}
\end{figure}

It is instructive to re-write Eq.(\ref{eq:hardest}) in such a way as to make direct contact with the corresponding Feynman diagrams:\\
\newcommand\hf{\textstyle{\frac12}}
\scalebox{0.74}{\begin{minipage}{1\textwidth}
\begin{eqnarray}
  \sigma_{1,\mbox{\tiny out=hardest}} &=& 2^5\left(-\frac{2\alpha_s}{\pi}\int_{Q_0}^Q\frac{dk_T}{k_T}\right)
  \left(\ln\frac{Q}{k_T}\right)
  \left(-\frac{2\alpha_s}{\pi}\int_{Q_0}^{k_T}\frac{dk_T'}{k_T'}\right)^3
\nonumber\\&& \hspace*{-2cm}
  \Biggl\langle\!m_0\Biggr|
  \Bigg[
    \frac{1}{3!}(-\hf Y\mathbf{t}_1\cdot\mathbf{t}_4)\,(+\hf i\pi\mathbf{t}_1\cdot\mathbf{t}_2)\,
      (+\hf i\pi\mathbf{t}_1\cdot\mathbf{t}_2)     
    +\frac{1}{3!}(+\hf i\pi\mathbf{t}_1\cdot\mathbf{t}_2)\,(+\hf i\pi\mathbf{t}_1\cdot\mathbf{t}_2)\,
      (-\hf Y\mathbf{t}_1\cdot\mathbf{t}_4)  \nonumber \\ & & \hspace*{0cm} 
    +\frac{1}{2!}\,(-\hf i\pi\mathbf{t}_1\cdot\mathbf{t}_2)\,(-\hf Y\mathbf{t}_1\cdot\mathbf{t}_4)\,
      (+\hf i\pi\mathbf{t}_1\cdot\mathbf{t}_2)  
    +\frac{1}{3!}\,(+\hf i\pi\mathbf{t}_1\cdot\mathbf{t}_2)\,(-\hf Y\mathbf{t}_1\cdot\mathbf{t}_4)\,
      (+\hf i\pi\mathbf{t}_1\cdot\mathbf{t}_2)            
  \Bigg]
  2(\hf\mathbf{t}_1^2)
\nonumber\\&& \hspace*{-1cm}
  - 2(\hf\mathbf{t}_1^{a\dagger})
  \Bigg[
    \frac{1}{3!}(-\hf Y\mathbf{T}_1\cdot\mathbf{t}_4)\,(+\hf i\pi\mathbf{T}_1\cdot\mathbf{T}_2)\,
      (+\hf i\pi\mathbf{T}_1\cdot\mathbf{T}_2)     
    +\frac{1}{3!}(+\hf i\pi\mathbf{T}_1\cdot\mathbf{T}_2)\,(+\hf i\pi\mathbf{T}_1\cdot\mathbf{T}_2)\,
      (-\hf Y\mathbf{T}_1\cdot\mathbf{T}_4)  \nonumber \\ & &
    +\frac{1}{2!}\,(-\hf i\pi\mathbf{T}_1\cdot\mathbf{T}_2)\,(-\hf Y\mathbf{T}_1\cdot\mathbf{T}_4)\,
      (+\hf i\pi\mathbf{T}_1\cdot\mathbf{T}_2)  
    +\frac{1}{3!}\,(+\hf i\pi\mathbf{T}_1\cdot\mathbf{T}_2)\,(-\hf Y\mathbf{T}_1\cdot\mathbf{T}_4)\,
      (+\hf i\pi\mathbf{T}_1\cdot\mathbf{T}_2)            
  \Bigg]  
  \mathbf{t}_1^a
  \Biggl|m_0\!\Biggr\rangle.
\nonumber \\ && \nonumber
\end{eqnarray}
\end{minipage}} \\
The eight terms in the matrix element correspond, in Feynman gauge, to the eight diagrams in Figure \ref{fig:hardest}, in the order (a), (g), (e), (c), (b), (h), (f), (d). The super-leading contribution arises when the soft gluons linking partons 1 and 2 are Coulomb gluons and the soft gluon shown linking partons 1 and 4 is an eikonal gluon. The hard gluons responsible for producing the hard scatter are represented by traditional curly lines. All the factors in this equation are understandable.
The signs on $\alpha_s$ and all the $\mathbf{t}_i\cdot\mathbf{t}_j$ terms
are the natural ones defined by our notation\footnote{Here and in the remainder of this paragraph we do not distinguish between $\mathbf{T}_i$ and $\mathbf{t}_i$.}.
  Of the numerical factors, those on $\alpha_s$ and the
$\mathbf{t}_i\cdot\mathbf{t}_j$ terms are again defined by our notation.
The $3!$ arises when all three gluons are on the same side of the cut and accounts for the strong ordering in transverse momentum. Similarly the $2!$ in the middle term arises when two of the three gluons are on the same side of the cut.  Of the five powers of two, two come from the
fact that $i \pi \mathbf{t}_1\cdot\mathbf{t}_2$ is shorthand for the sum of
all Coulomb diagrams, one comes from the fact that
$-Y\mathbf{t}_1\cdot\mathbf{t}_4$ is shorthand for the sum over all gluons
exchanged across the gap, one comes from accounting for the hermitian conjugate amplitudes and the final factor comes from allowing the out-of-gap gluon to be
collinear to either of the incoming quarks\footnote{Strictly speaking
this should be performed by summing over exchange of the identities of
partons~1 and~2 when they are different, but since it turns out that the
final result is symmetric in partons~1 and~2, simply multiplying by~2
suffices.}.
The $\hf\mathbf{t}_1^2$ term (and the corresponding $\hf\mathbf{t}_1^{a \dagger} \cdots \mathbf{t}_1^a$ term in the real emission case) deserves a little discussion. It is the colour factor corresponding to the out-of-gap gluon and as such it follows (in the collinear limit) after summing over graphs where a gluon connects parton 1 with partons 2, 3 or 4. Each of these attachments is illustrated by a different dotted line in the figures. The sum over diagrams leads to a colour diagonal contribution (a result that is more transparent in a physical gauge).
The extra factor of 2 in front of the $\hf \mathbf{t}_1^2$ term comes
from the fact that the virtual, out-of-gap, gluon can be either side of
the cut, i.e. in Figure~\ref{eq:hardest} the $\mathbf{t}_1^2$ gluon
could also appear to the right of the cut in each diagram. Similarly the
real gluon could also be attached to parton 1 on the right of the cut
and any other parton on the left of the cut. Strictly speaking this doubles the total number of graphs; although we do not show the additional $3 \times 8$ graphs (because they produce identical results in the collinear limit) they are distinct from the $3 \times 8$ shown in Figure \ref{fig:hardest}.

To complete the calculation, we need to compute the contribution arising
from the case where there is one virtual emission of higher $k_T$ than the
out-of-gap emission. Now we use the expression for  $\{~\}_2$ derived in Eq.(\ref{eq:a2}) in conjunction
with the order $\alpha_s$ expansion of the exponential factors that lie outside of the main bracket in 
Eq.~(\ref{fullresult}). Note that this is the only remaining contribution to the lowest order
super-leading logarithm since all lower order expansions of the main bracket in Eq.~(\ref{fullresult}) 
(i.e. $\{ ~ \}_1$ and $\{ ~ \}_0$) vanish identically. The result is
\begin{eqnarray}
  \sigma_{1,\mbox{\tiny out=second-hardest}} &=& \left(-\frac{2\alpha_s}{\pi}\right)^4 \int_{Q_0}^Q\frac{dk_T}{k_T}
  \left(2\ln\frac{Q}{k_T}\right)
 \left( \int_{k_T}^{Q}\frac{dk_T'}{k_T'} \right) 
  \left(\int_{Q_0}^{k_T}\frac{dk_T''}{k_T''}\right)^2 \nonumber \\ & & \hspace*{-3cm} \frac{i\pi Y}{2} 
   (4 \pi i) 
   \Biggl\langle\!m_0\Biggr|
  (\mathbf{t}_1 \cdot \mathbf{t}_2)
  \left(
  \mathbf{t}_1^2
   [\mathbf{t}_1\cdot\mathbf{t}_4,\mathbf{t}_1\cdot\mathbf{t}_2]
  -\mathbf{t}_1^{a\dagger}
    [\mathbf{T}_1\cdot\mathbf{T}_4,\mathbf{T}_1\cdot\mathbf{T}_2 ] \mathbf{t}_1^a \right) \Biggl|m_0\!\Biggr\rangle.
\label{eq:nexthardest}
\end{eqnarray}
The factor $4 \pi i$ contains a factor of 2 for the contribution where the hardest Coulomb gluon is on the left of the cut, i.e. the $(\mathbf{t}_1 \cdot \mathbf{t}_2)$ factor on the far left of the matrix element is moved to the far right.
Evaluating the colour matrix elements and performing the integrals
over transverse momenta we confirm the result quoted in Eq.(\ref{eq:resnexthardest}).

\begin{figure}[t]
\centering
\subfigure[]{\includegraphics[width=0.4\textwidth]{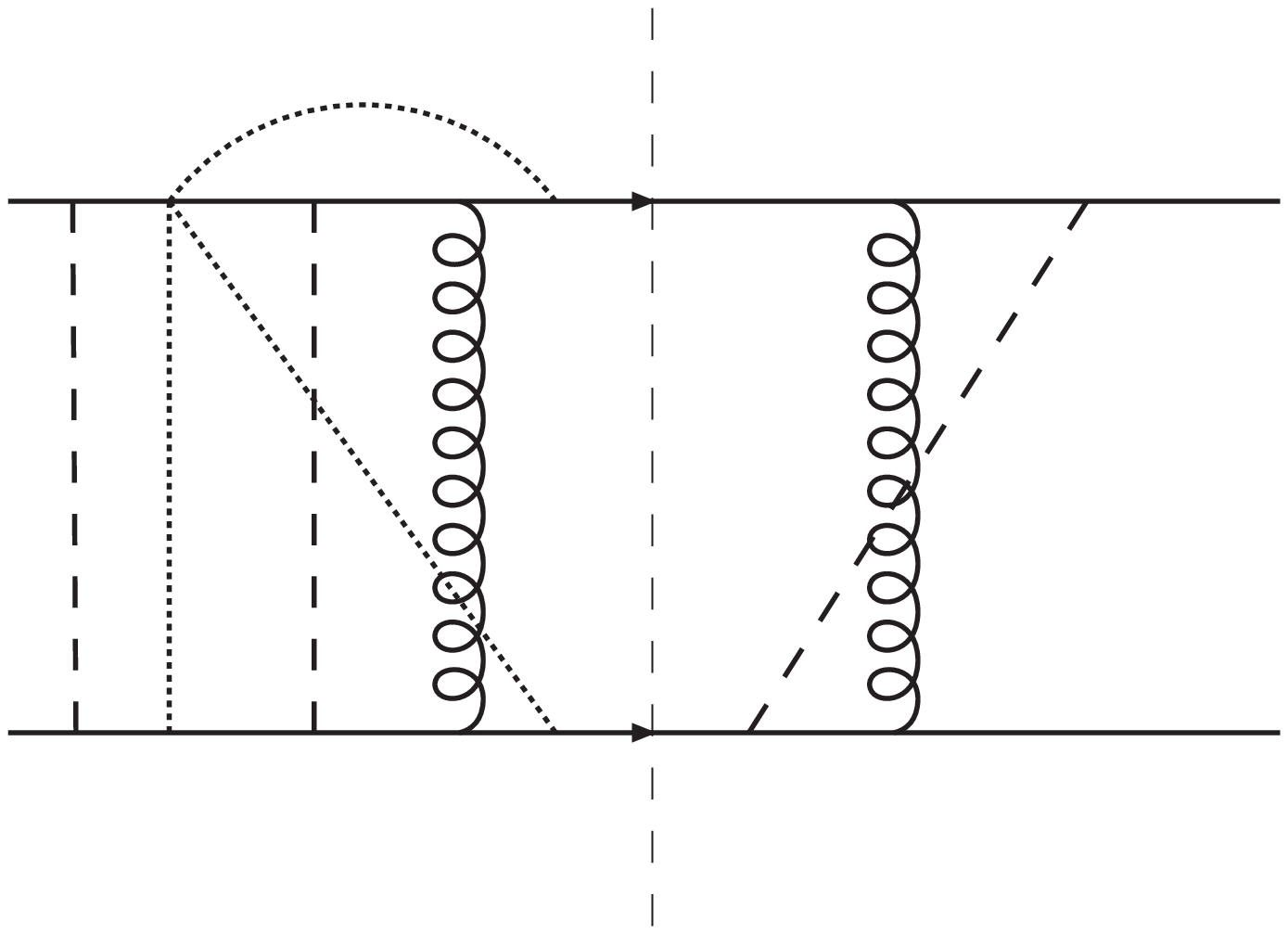}}
\subfigure[]{\includegraphics[width=0.4\textwidth]{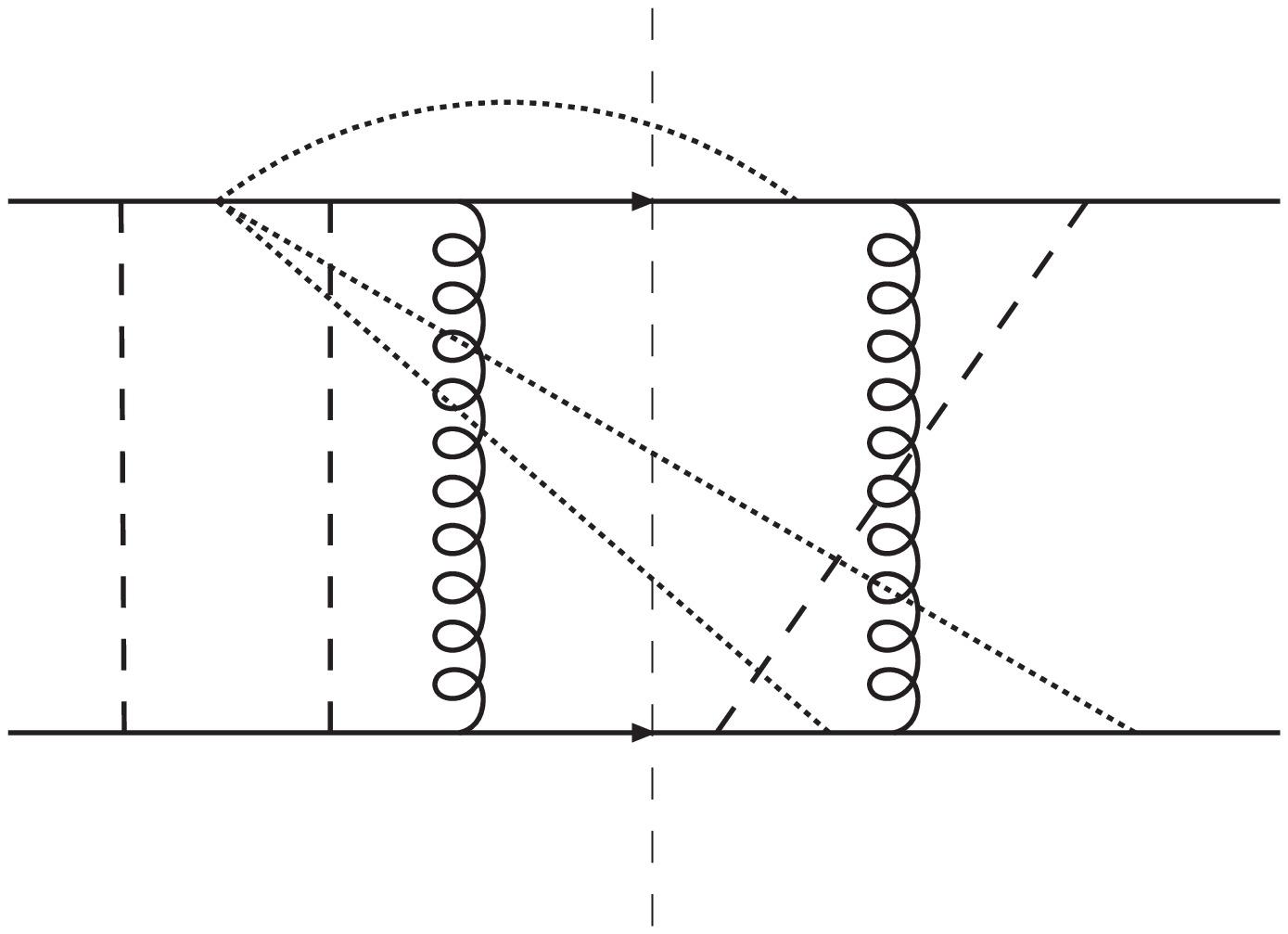}} \\
\subfigure[]{\includegraphics[width=0.4\textwidth]{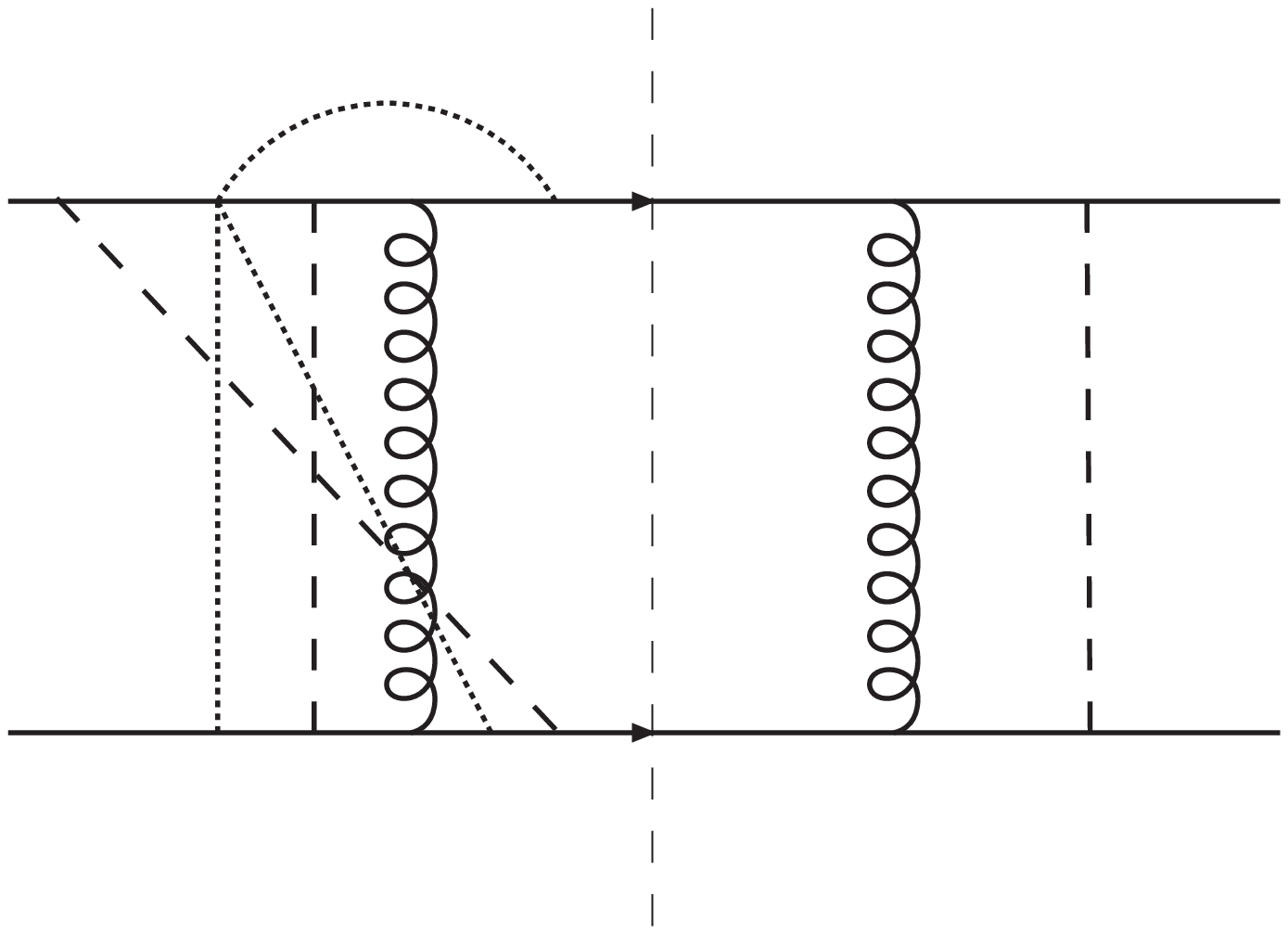}}
\subfigure[]{\includegraphics[width=0.4\textwidth]{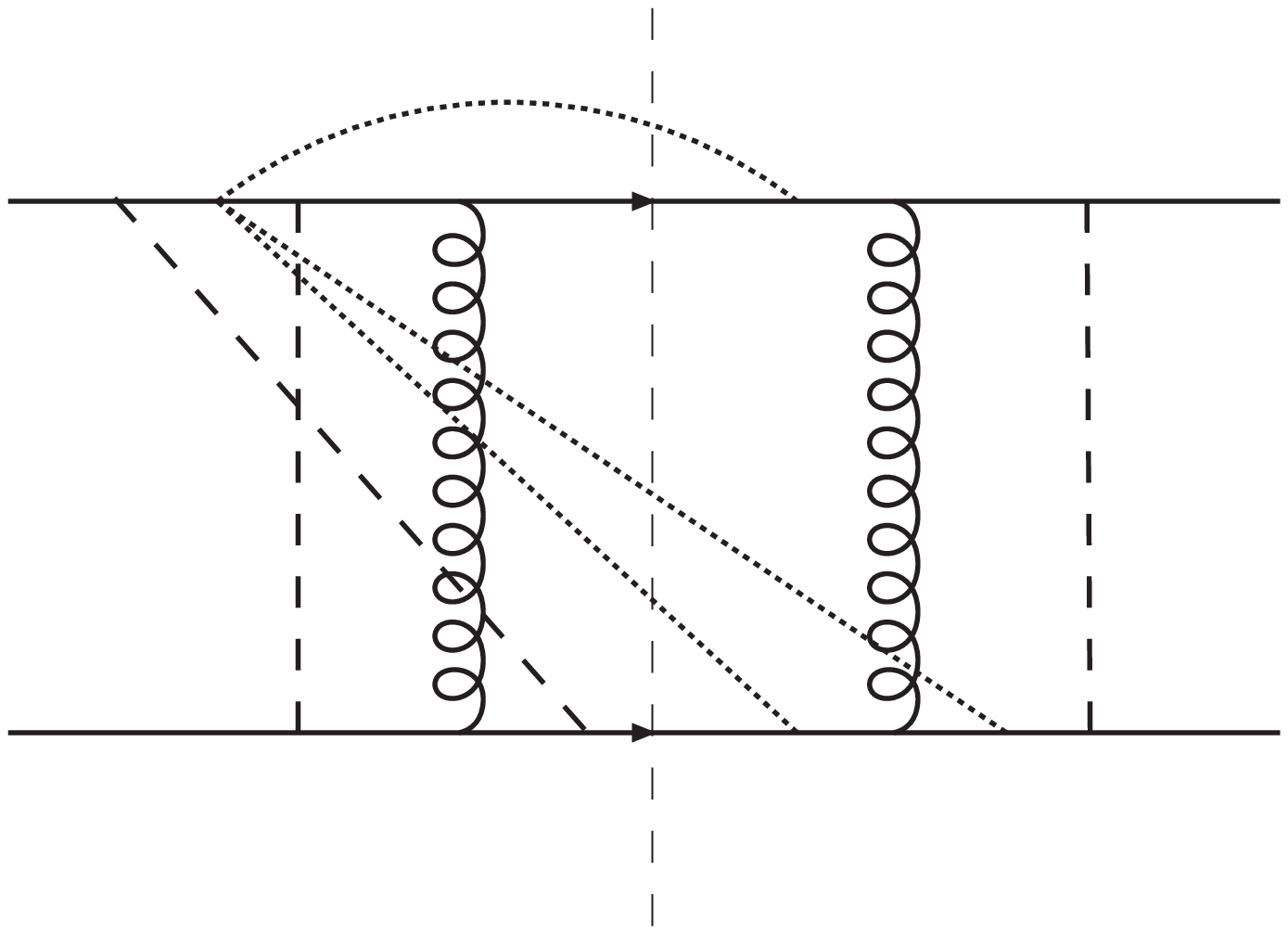}} 
\caption{The relevant Feynman diagrams in the case that the out-of-gap (dotted) gluon is the second hardest gluon. The dashed lines indicate soft (eikonal and Coulomb) gluons. Each subfigure represents three Feynman diagrams, corresponding to the three different ways of attaching the out-of-gap gluon.
The soft gluon to the right of the cut should only be integrated over the
region in which it has transverse momentum less than the out-of-gap
gluon.}
\label{fig:nexttohardest}
\end{figure}

Again we can do a Feynman diagram decomposition, see Figure \ref{fig:nexttohardest}:\\
\scalebox{0.92}{\begin{minipage}{1\textwidth}
\begin{eqnarray}
  \sigma_{1,\mbox{\tiny out=second-hardest}} &=& 2^5\left(-\frac{2\alpha_s}{\pi} \right)^4 \int_{Q_0}^Q\frac{dk_T}{k_T}   \left(\ln\frac{Q}{k_T}\right)
  \int_{k_T}^{Q}\frac{dk_T'}{k_T'} \left( \int_{Q_0}^{k_T}\frac{dk_T''}{k_T''}\right)^2
\nonumber\\&& \hspace*{-2.5cm}
  \Biggl\langle\!m_0\Biggr|
  \Bigl(
   (-\hf Y\mathbf{t}_1\cdot\mathbf{t}_4)\,(+\hf i\pi\mathbf{t}_1\cdot\mathbf{t}_2) 
    +(-\hf i\pi\mathbf{t}_1\cdot\mathbf{t}_2)\,(-\hf Y\mathbf{t}_1\cdot\mathbf{t}_4)
  \Bigr)
  2(\hf\mathbf{t}_1^2)
  (+\hf i\pi \mathbf{t}_1 \cdot \mathbf{t}_2)
\nonumber\\&& \hspace*{-2.5cm}
  -2(\hf\mathbf{t}_1^{a\dagger})
  \Bigl(
    (-\hf Y\mathbf{T}_1\cdot\mathbf{T}_4)\,(+\hf i\pi\mathbf{T}_1\cdot\mathbf{T}_2) 
    +(-\hf i\pi\mathbf{T}_1\cdot\mathbf{T}_2)\,(-\hf Y\mathbf{T}_1\cdot\mathbf{T}_4)
  \Bigr)
   \mathbf{t}_1^a (+\hf i\pi \mathbf{t}_1 \cdot \mathbf{t}_2)
  \Biggl|m_0\!\Biggr\rangle.
\nonumber \\ && \nonumber
\end{eqnarray}
\end{minipage}}\\
The first two terms in the matrix element correspond to graphs (a) and (c) whilst the latter pair correspond to graphs (b) and (d). Again $\mathbf{t}_1^2$ (and $\mathbf{t}_1^{a \dagger} \cdots \mathbf{t}_1^a$) is shorthand for the sum over graphs where the out-of-gap real gluon connects parton 1 with partons 2, 3 and 4, and we do not show the additional $3 \times 4$ graphs that occur when the out-of-gap gluon couples to parton 1 on the right of the cut. 

\section{Results}
\label{results}

Our main results are Eqs.~(\ref{eq:hardest}) and~(\ref{eq:nexthardest}),
or equivalently the Feynman diagrams in Figures~\ref{fig:hardest}
and~\ref{fig:nexttohardest}.  In the case that the scattered particles
are quarks, one can check that they reproduce the results given in
\cite{Forshaw:2006fk} and Section~\ref{summary},
\begin{eqnarray}
  \sigma_{1,\mbox{\tiny out=hardest},qq} &=&
  -\sigma_0\left(\frac{2\alpha_s}\pi\right)^4
  \ln^5\left(\frac{Q}{Q_0}\right)\pi^2Y\frac{N^2-2}{240}\,, \\
  \sigma_{1,\mbox{\tiny out=second-hardest},qq} &=&
  -\sigma_0\left(\frac{2\alpha_s}\pi\right)^4
  \ln^5\left(\frac{Q}{Q_0}\right)\pi^2Y\frac{N^2-1}{120}\,, \\
  \sigma_{1,qq} &=&
  -\sigma_0\left(\frac{2\alpha_s}\pi\right)^4
  \ln^5\left(\frac{Q}{Q_0}\right)\pi^2Y\frac{3N^2-4}{240}\,.
\end{eqnarray}
The simplicity of the results in colour basis independent
notation also allows us to calculate the coefficient of the super-leading
logarithm for other scattering processes, in particular involving
gluons. The colour algebra is no more complicated in principle than for quarks,
and we have checked our results using the \textsf{Colour} program\cite{Hakkinen:1996bb}.
For quark--gluon scattering we obtain
\begin{eqnarray}
  \sigma_{1,\mbox{\tiny out=hardest},qg} &=&
  -\sigma_0\left(\frac{2\alpha_s}\pi\right)^4
  \ln^5\left(\frac{Q}{Q_0}\right)\pi^2Y\frac{N^2}{80}\,, \\
  \sigma_{1,\mbox{\tiny out=second-hardest},qg} &=&
  -\sigma_0\left(\frac{2\alpha_s}\pi\right)^4
  \ln^5\left(\frac{Q}{Q_0}\right)\pi^2Y\frac{N^2}{60}\,, \\
  \sigma_{1,qg} &=&
  -\sigma_0\left(\frac{2\alpha_s}\pi\right)^4
  \ln^5\left(\frac{Q}{Q_0}\right)\pi^2Y\frac{7N^2}{240}\,.
\end{eqnarray}
This result is remarkable in that it does not depend upon whether the out-of-gap gluon is collinear to the quark or to the gluon. For gluon--gluon scattering we find
\begin{eqnarray}
  \sigma_{1,\mbox{\tiny out=hardest},gg} &=&
  -\sigma_0\left(\frac{2\alpha_s}\pi\right)^4
  \ln^5\left(\frac{Q}{Q_0}\right)\pi^2Y\frac{5N^2+12}{240}\,, \\
  \sigma_{1,\mbox{\tiny out=second-hardest},gg} &=&
  -\sigma_0\left(\frac{2\alpha_s}\pi\right)^4
  \ln^5\left(\frac{Q}{Q_0}\right)\pi^2Y\frac{N^2}{60}\,, \\
  \sigma_{1,gg} &=&
  -\sigma_0\left(\frac{2\alpha_s}\pi\right)^4
  \ln^5\left(\frac{Q}{Q_0}\right)\pi^2Y\frac{3N^2+4}{80}\,.
\end{eqnarray}
Note that in all cases $\sigma_0$ is the corresponding Born level cross-section (i.e. it differs by a colour factor for each sub-process). The results for processes involving anti-quarks are identical to the corresponding processes involving quarks.

\section{Outlook}
\label{outlook}

\begin{figure}[t]
\centering
\includegraphics[width=0.8\textwidth]{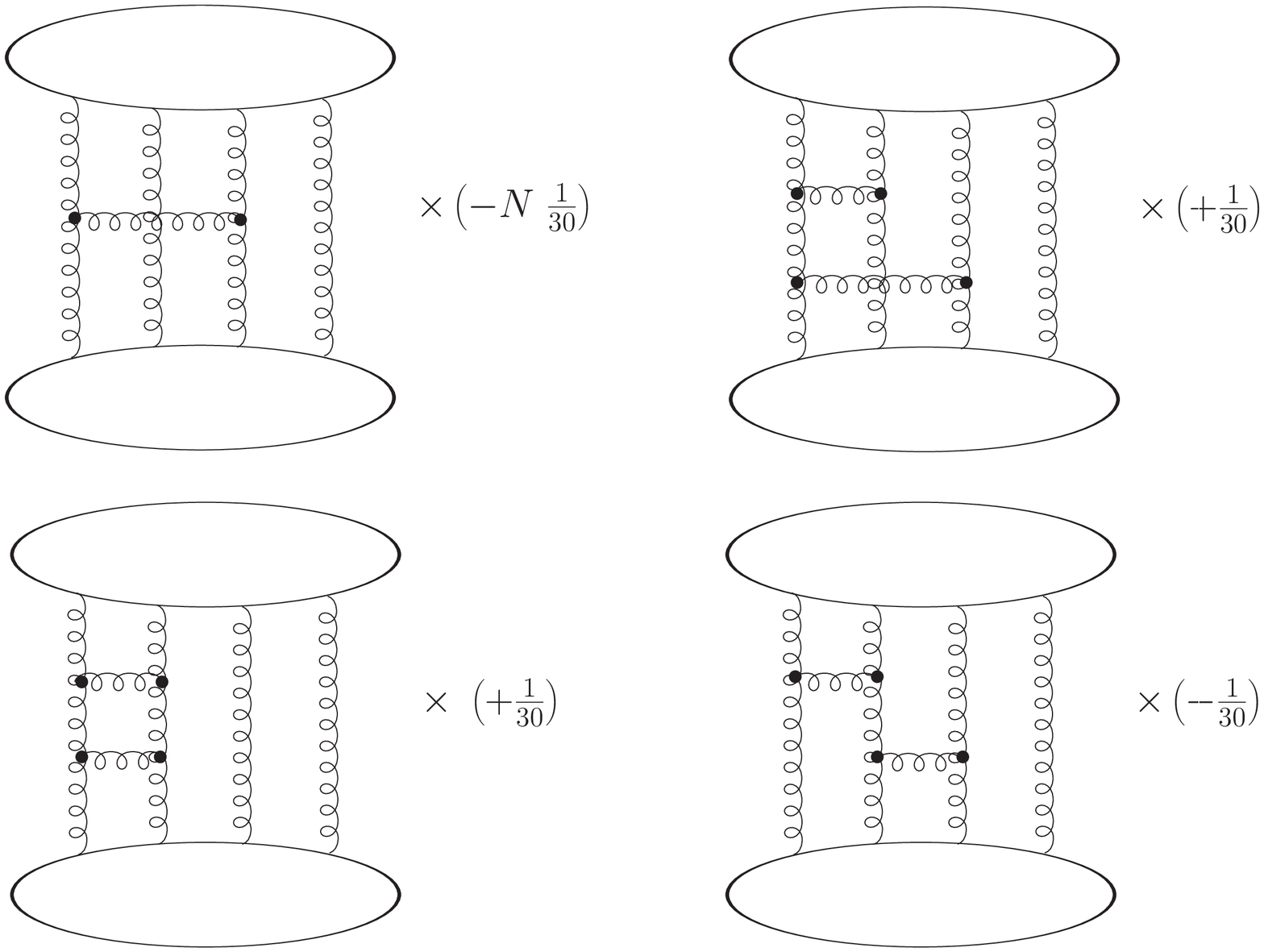}
\caption{The four diagrams that generate the colour matrix elements
  when the out-of-gap gluon is the hardest gluon. In the case that the
out-of-gap gluon is next-to-hardest, only the first diagram
contributes. The upper and lower loops can be quarks, anti-quarks or
gluons. Note that these are not Feynman diagrams or even uncut
diagrams; they represent only the colour factor of the final result.
In the first diagram one of the original six gluon lines has been contracted away, resulting in the additional factor of $N_c$.}
\label{fig:traces}
\end{figure}

In this paper we have reconsidered the super-leading logarithms
discovered for gaps between jets observables in
\cite{Forshaw:2006fk} using a colour basis independent notation.
It has added considerable insight, allowing their calculation to be
condensed down to a small number of Feynman diagrams
(Figures~\ref{fig:hardest} and~\ref{fig:nexttohardest}) and allowing the
first super-leading logarithm to be calculated for gluon scattering
processes.  The main questions still remain open however: the structure
of higher order super-leading logarithms; how widespread they are in
other observables for hadron collisions; and whether they can be
reorganized and resummed or removed by a suitable redefinition of
observables or of incoming partonic states.  We hope that the insight
provided by the colour basis independent notation will ultimately help
to illuminate these questions.

We close by noting that further simplifications are possible. In particular, the
commutators between gluon exchanges from an external leg in different
orders can be written as emission off the exchanged Coulomb gluons. The
colour matrix elements in Eqs.~(\ref{eq:hardest}) and~(\ref{eq:nexthardest}) 
can then be shown to be identical to those illustrated in Figure
\ref{fig:traces}. In this form it is clear that the superleading logarithms
arise as a result of the non-Abelian nature of the Coulomb gluon
interaction. It is also now clear why the coefficient of the
superleading logarithm is independent of whether the out-of-gap gluon
is collinear to parton~1 or~2: the result is invariant under
interchange of the particle types in the upper and lower loops.

\section*{Acknowledgements}

We are grateful to Stefano Catani for suggesting the colour basis
independent notation used here.  This work was begun at the Galileo
Galilei Institute for Theoretical Physics workshop entitled ``Advancing
Collider Physics: from Twistors to Monte Carlos''; MHS gratefully
acknowledges the Institute's financial support.  We also thank Mrinal
Dasgupta, James Keates, Malin Sj\"odahl and George Sterman for interesting discussions of these and
related topics.

\end{document}